\documentclass[format=acmsmall, review=false, screen=true]{acmart}

\usepackage{graphicx}
\usepackage{caption} 
\usepackage{array}
\usepackage{comment}
\usepackage{pdflscape}
\usepackage{threeparttable}
\usepackage{longtable}
\setcitestyle{square}

\usepackage{arydshln}

\AtBeginDocument{%
  \providecommand\BibTeX{{%
    \normalfont B\kern-0.5em{\scshape i\kern-0.25em b}\kern-0.8em\TeX}}}

\setcopyright{acmcopyright}

\acmJournal{CSUR}

\begin{document}

\title{Application Management in Fog Computing Environments: A Taxonomy, Review and Future Directions}
\author{Redowan Mahmud}
\orcid{ 0000-0003-0785-0457}
\author{Kotagiri Ramamohanarao}
\author{Rajkumar Buyya}
\affiliation{%
  \institution{\\University of Melbourne}
  \department{Cloud Computing and Distributed Systems (CLOUDS) Laboratory, School of Computing and Information Systems}
  \streetaddress{Parkville Campus}
  \city{Melbourne}
  \state{Victoria}
  \postcode{3010}
  \country{Australia}}
\email{mahmudm@student.unimelb.edu.au}

\renewcommand{\shortauthors}{R. Mahmud et al.}

\begin{abstract}
The Internet of Things (IoT) paradigm is being rapidly adopted for the creation of smart environments in various domains. The IoT-enabled Cyber-Physical Systems (CPSs) associated with smart city, healthcare, Industry 4.0 and Agtech handle a huge volume of data and require data processing services from different types of applications in real-time. The Cloud-centric execution of IoT applications barely meets such requirements as the Cloud datacentres reside at a multi-hop distance from the IoT devices. \textit{Fog computing}, an extension of Cloud at the edge network, can execute these applications closer to data sources. Thus, Fog computing can improve application service delivery time and resist network congestion. However, the Fog nodes are highly distributed, heterogeneous and most of them are constrained in resources and spatial sharing. Therefore, efficient management of applications is necessary to fully exploit the capabilities of Fog nodes. In this work, we investigate the existing application management strategies in Fog computing and review them in terms of architecture, placement and maintenance. Additionally, we propose a comprehensive taxonomy and highlight the research gaps in Fog-based application management. We also discuss a perspective model and provide future research directions for further improvement of application management in Fog computing.  
\end{abstract}


\maketitle

\section{Introduction}
The Internet of Things (IoT) concept has changed the structure of material environments by connecting numerous computing components, digital machines, dumb objects and animals with the Internet. IoT enables them to perceive the external ambiance as sensors and trigger any action based on the given commands using actuators \cite{Gubbi}. Thus, IoT creates a novel type of interaction among different real-world entities in ingenious ways. The ongoing advancement in the field of hardware and communication technology is consistently improving and expanding the applicability of IoT that consequently helps in realizing the theory of smart city, remote healthcare, Industry 4.0 and Agtech \cite{Interoperability}. Recently, various Cyber-Physical Systems (CPS) for smart environments such as indoor locator, digital health recorder and robot-assisted supply chain manager have been developed through the widespread deployment of IoT devices. Moreover, according to the current trend of practising IoT, it is expected that by 2030, there will be 1.2 trillion active IoT devices with potential economic impact of \$15 trillion per year \cite{forecast}. 
\par IoT devices can generate data incessantly or periodically. These data are filtered, analysed and evaluated by different types of applications \cite{IoTapp}. As most of the IoT devices are equipped with limited energy, computing and networking capabilities, they are considered ill-suited to execute heavyweight applications \cite{IoTPower}. Moreover, based on the working environments of IoT-enabled CPSs, corresponding applications often require data processing within a defined time frame. Their data-driven interactions with the IoT devices also demands less-congested network. The computing infrastructure executing the applications for IoT-enabled CPSs needs to observe these issues so that the desired responsiveness of the CPSs can be ensured \cite{Ritu-ICSOC}. 
\subsection{Scope and Benefits of Fog Computing}
Cloud computing has been the backbone for hosting and offering subscription-oriented computing and application services. It is also used to execute the applications for different IoT-enabled CPSs \cite{CloudIoT}. The Cloud datacentres consist of data and computing servers to facilitate users with storage and virtualized computing instances \cite{Buyya}. Nevertheless, these datacentres are located at a multi-hop distance from the IoT devices. Therefore, a longer period of time is required to transfer data and command between the IoT devices and the applications executing on the Cloud instances that also degrades the application service delivery time \cite{Realtime}. Furthermore, when a large number of IoT devices initiate data-driven interactions with remote applications, a substantial load is added to the network and severe congestion occurs. The computational overhead on Cloud datacentres also increases \cite{Fog1}. Because of these limitations, Cloud-centric application execution model often fails to meet the service requirements of different IoT-enabled CPSs. To address them, an extension of Cloud computing named \textit{Fog computing} was introduced by CISCO in 2012 \cite{Bonomi}.          
\begin{figure}[!t]
\centering 
\includegraphics[width=140mm, height= 75mm]{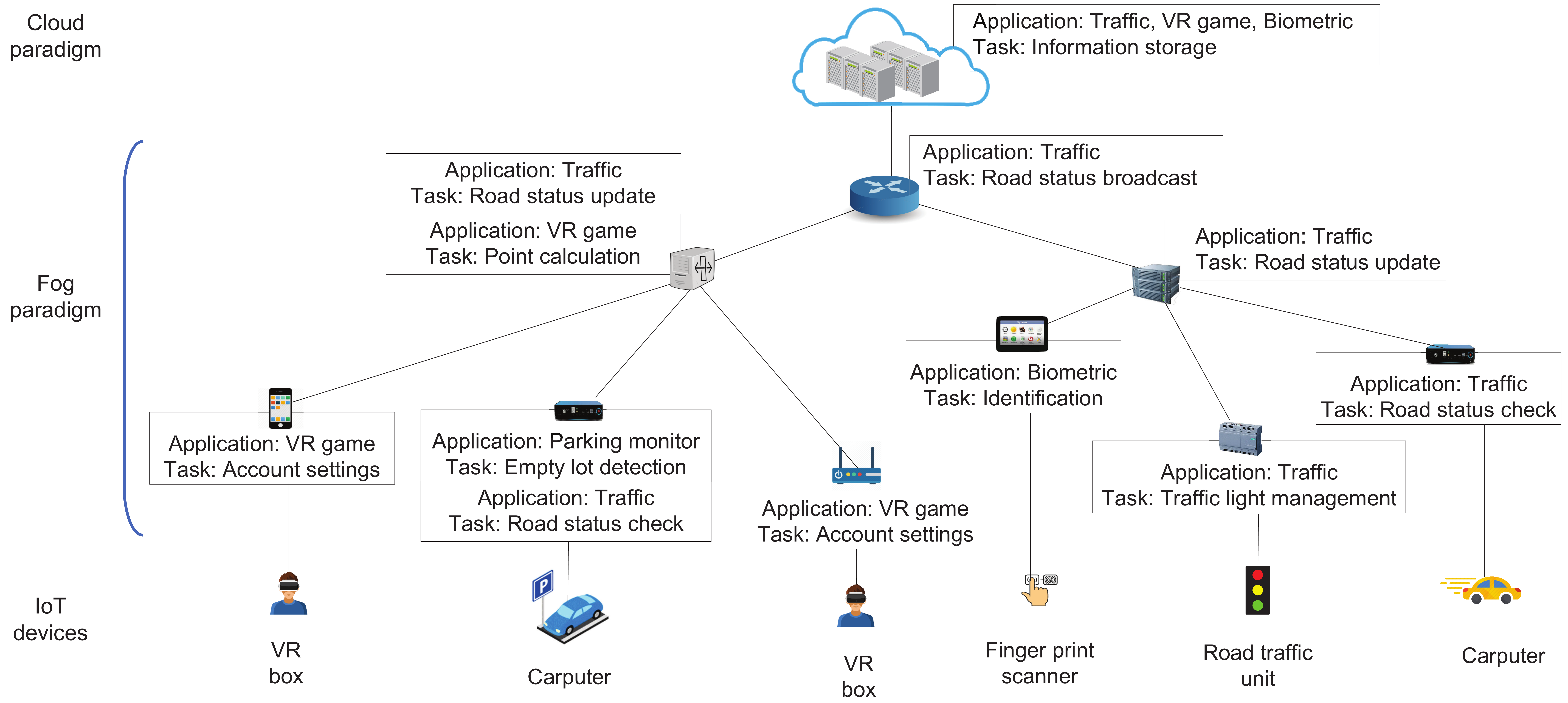}
\caption{An illustration of application execution in Fog environments}
\label{Fig:basic}
\end{figure}
\par Fog manages an intermediate layer between end user devices and Cloud datacentres by utilizing the computing components within the edge network \cite{Yousefpour}. In Fog environments, these computing components; for example personal computers, gateways, Raspberry PIs, nano-servers and micro-datacentres, are commonly known as \textit{Fog nodes}. As shown in Fig. \ref{Fig:basic}, the Fog nodes execute various IoT applications in proximity of data sources. Hence, Fog computing resists a huge amount of data from sending towards Cloud datacentres and decreases the data propagation delay. Consequently, the service latency of different applications improves \cite{100}. Moreover, Fog computing conserves network bandwidth that reduces the scope of network congestion. Through Fog computing, providers migrate the computational load from Cloud datacentres to network edge. As Fog nodes are less expensive, Fog computing lowers the operational cost of providers, saves energy for the Cloud datacentres, and improves the Quality of Experience (QoE) of users. Additionally, Fog supports robust location-awareness to simplify the communication with mobile and energy constrained end user devices \cite{Puliafito}. 
\par On the basis of the aforementioned features, Fog computing is considered very promising compared to Cloud in meeting the application service requirements of different IoT-enabled CPSs. Therefore, several technology giants such as Amazon, Alphabet and Microsoft have already started integrating Fog computing with their Cloud infrastructure \cite{big3}. Also there are other companies like Sonm, NEC Laboratories, FogHorn Systems and Drofika Labs that intend to make software systems for Fog environments \cite{FogHorn}. The development of FogFlow framework is regarded as a successful attempt to this direction \cite{FogFlow}. There also exists a Fog-based real-time patient monitoring system developed by Tata Consultancy Services \cite{Tata}.
\subsection{Application Management in Fog Computing}
Fog computing creates a wide distribution of infrastructure and platform services. Infrastructure services include on-demand exploitation of computing, networking (bandwidth and firewalls) and storage resources, whereas platform services facilitate application runtime environments, operating systems and programming interfaces \cite{Intro-Fog_Computing_principles}. Fog resource management denotes the administrative operations such as deployment, virtualization and monitoring of Fog nodes that foster the Fog-based infrastructure and platform services \cite{Intro-Assessment_Suitability}. Additionally, Fog resource management realizes load balancing, dynamic provisioning and auto-scaling to ensure service availability and multi-tenancy \cite{Hong}.  
\par Efficient Fog resource management assists IoT-enabled CPSs to operate multiple applications simultaneously. However, the characteristics of these applications vary from one CPS to another. For example, the expected application service delivery time for a CPS that remotely monitors the respiratory functions of critical patients is quite stringent compared to a CPS which measures the environmental parameters \cite{FogBus}. Moreover, an application that assists a CPS to perform virtual reality operations handles huge amount of data in per unit time compared to an application which helps in tracking the empty parking slots. Such diversified characteristics play vital roles in defining the Quality of Service (QoS) requirements of the applications that cannot be met only through Fog resource management \cite{Edge}. This perception also urges to develop different application management strategies according to the preferences of the applications. Usually, an application management strategy refers to a collection of algorithms, mathematical models, empirical analysis and recommendations that regulate the implementation, installation and execution of applications in a computing environment. Moreover, application management strategies practice admission control, location transparency, data maintenance and service resiliency as per the demands of the corresponding system \cite{Demystify}. Nevertheless, there are three research questions that become crucial while developing application management strategies for Fog computing environments. They are discussed below: 
\begin{itemize}
\item \textit{How should the applications be composed?}
\par To address this question, an application management strategy requires to specify the features of applications such as their programming model, functional layout, service type, workload type so that they can be aligned with the Fog-based infrastructure and platform services.
\item \textit{How should the applications be placed?}
\par To address this question, an application management strategy requires to find suitable placement options for the applications in Fog environments. At the same time, the strategy needs to make a balance between application-centric QoS requirements.    
\item \textit{How should the applications be maintained?}
\par To address this question, an application management strategy requires to facilitate security and resiliency support during application execution in Fog environments. Moreover, the strategy needs to monitor the performance of the applications in consistent manner.          
\end{itemize} 
\par Although operational responses to these questions are important for efficient application management, their actual realization in Fog computing is a very challenging task.   
\subsubsection{Challenges of Application Management:}
The challenges faced by application management strategies in Fog environments are listed as: 
\par $\bullet$ \textit{Resource and energy constrained, distributed and heterogeneous Fog nodes}: Most of the Fog nodes are constrained in processing power, networking capability, storage and energy capacity. They are deployed in distributed order at the edge network. Their communication standards and operating systems also vary from one to another \cite{Fog1}. As a consequence, the time-optimized and platform independent application management become tedious to ensure in Fog. Additionally, Fog infrastructure is less flexible than Cloud in terms of sharing resources; for example, the Fog nodes located in California cannot be harnessed for the CPSs in Melbourne. Such constraints limit the execution domain for large-scale IoT applications in Fog \cite{50}.         
\par $\bullet$ \textit{Subjected to uncertain failures}: Fog nodes are highly prone to get affected by anomalies such as power failures and out of capacity faults that obstruct the execution of applications assigned to them \cite{62}. Due to latency issues, the recovery of applications also becomes difficult.
\par $\bullet$ \textit{Standard-less integration}: The applications executing in Fog often need the services offered by different computing paradigms. For example, the Fog-based health data analytic applications require the Cloud-based storage service to facilitate location-independent medical report sharing. In such scenarios, integration of Fog infrastructure with others is necessary \cite{35}. Nevertheless, the absence of efficient frameworks and standards resist the Fog environments to provide this assistance to the applications.    
\par $\bullet$ \textit{Lack of interoperability}: The structural differences between Fog and Cloud environments obstruct the interoperability of IoT applications. Due to lack of interoperability, an extensive programming effort is required to customize the existing Cloud-based IoT applications so that they can leverage the benefits of Fog computing \cite{Interoperability}.          
\par $\bullet$ \textit{Absence of business model}: Fog environments lacks business models that hampers the budget management of users and the profit enhancement of providers. These monetary issues consequently resist both the parties to execute applications in Fog \cite{49}.
\par $\bullet$ \textit{Inefficient task distribution}: Fog environments operate in decentralized manner across the edge network. The coexistence of multiple decision-making entities increases the application management complexity in Fog environments that ultimately results in poor distribution of application tasks over the Fog nodes \cite{Baccarelli}.   
\par $\bullet$ \textit{Less secured}: The outcomes of applications executing in Fog can be requested by different types of users. For example, the results of a Fog-based healthcare application are relevant to the hospitals, insurance companies and employer organizations. In such cases, despite of the necessity, the on-demand and secured access to application outcomes becomes difficult to ensure because of the resource scarcity and orientation of Fog environments \cite{Shirazi}.         
\subsubsection{Motivation of Research:}
\begin{figure}[!t]
\centering 
\captionsetup{justification=centering,margin=2cm}
\includegraphics[width=65mm, height= 20mm]{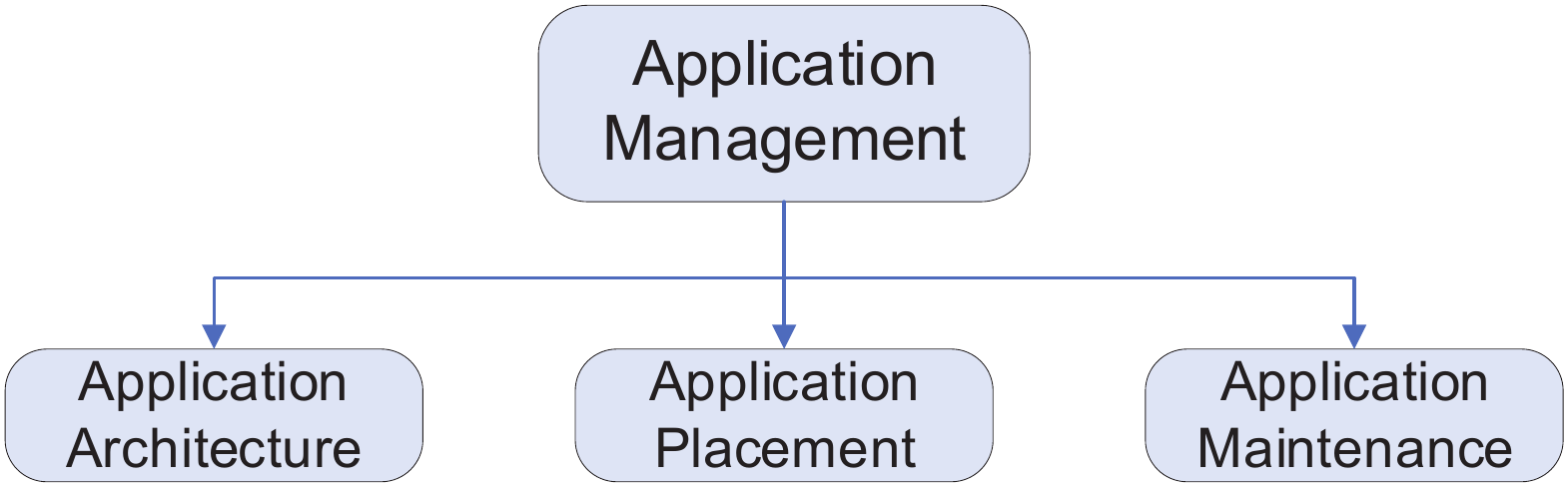}
\caption{Different aspects of application management}
\label{Fig:aspects}
\end{figure}

Considering the associated challenges, a notable number of application management strategies have already been developed for Fog computing environments. They predominantly focus on the modularization of applications to deal with the resource constraints of Fog nodes \cite{21} \cite{28} \cite{76}. These strategies also adopt the web service-based communication techniques to simplify the interactions between different components of modular applications hosted on distributed Fog nodes \cite{60} \cite{96}. While assigning the applications to the Fog nodes, the existing application management strategies give much emphasis on meeting the service delivery deadline and optimizing the cost and energy consumptions \cite{57} \cite{126} \cite{40}. Most of them operate discretely and apply strict synchronization measures over the Fog nodes to mitigate the effect of interference \cite{Profit} \cite{92}. The application management strategies also incorporate both proactive and reactive fault tolerance techniques to support the reliable execution of the applications in Fog environments \cite{9} \cite{30} \cite{73}. However, in the literature, very few initiatives have been found that categorize the application management strategies in a systematic way \cite{Yousefpour} \cite{Naha}. Therefore, in this work, we identify three important aspects of application management in Fog computing environments namely application architecture, application placement and application maintenance, as shown in Fig. \ref{Fig:aspects} and present separate taxonomy for each of them. Based on the proposed taxonomy, we also review the existing application management strategies and denote how the research community can leverage the solutions to make further progress. The \textbf{main contributions} of this work are: 
\begin{itemize}
\item We review the existing literature on application management strategies in Fog from the perspectives of architecture, placement and maintenance and propose their taxonomy. 
\item We discuss a framework that is logically distributed and helps adaptive and holistic management of applications in Fog computing environments  
\item We identify the research gaps in Fog computing-based application management and highlight future research directions for further improvement in this field.
\end{itemize}      
\subsection{Article Organization}
The rest of the article is organized as follows. The differences between Fog and other contemporary computing paradigms along with the description of related surveys are illustrated in Section \ref{Sec_Background}. In Section \ref{Sec_Design}, a discussion on application architecture in Fog environments is presented. Section \ref{Sec_Placement} highlights the existing techniques to place applications in Fog environments. Section \ref{Sec_Maintenance} explores the application maintenance operations. Section \ref{Sec_Model} demonstrates a perspective framework for Fog-based application management. Section \ref{Sec_Future} provides future direction to improve the application management strategies in Fog environments based on the identified research gaps in Section \ref{Sec_Placement}-\ref{Sec_Maintenance}. Finally, Section \ref{Sec_Conclusion} summarizes our efforts and concludes the survey. 
\section{Background Study}\label{Sec_Background}
\subsection{Comparison among Mist, Edge and Fog computing}
\begin{figure}[!t]
\centering 
\includegraphics[width=100mm, height= 60mm]{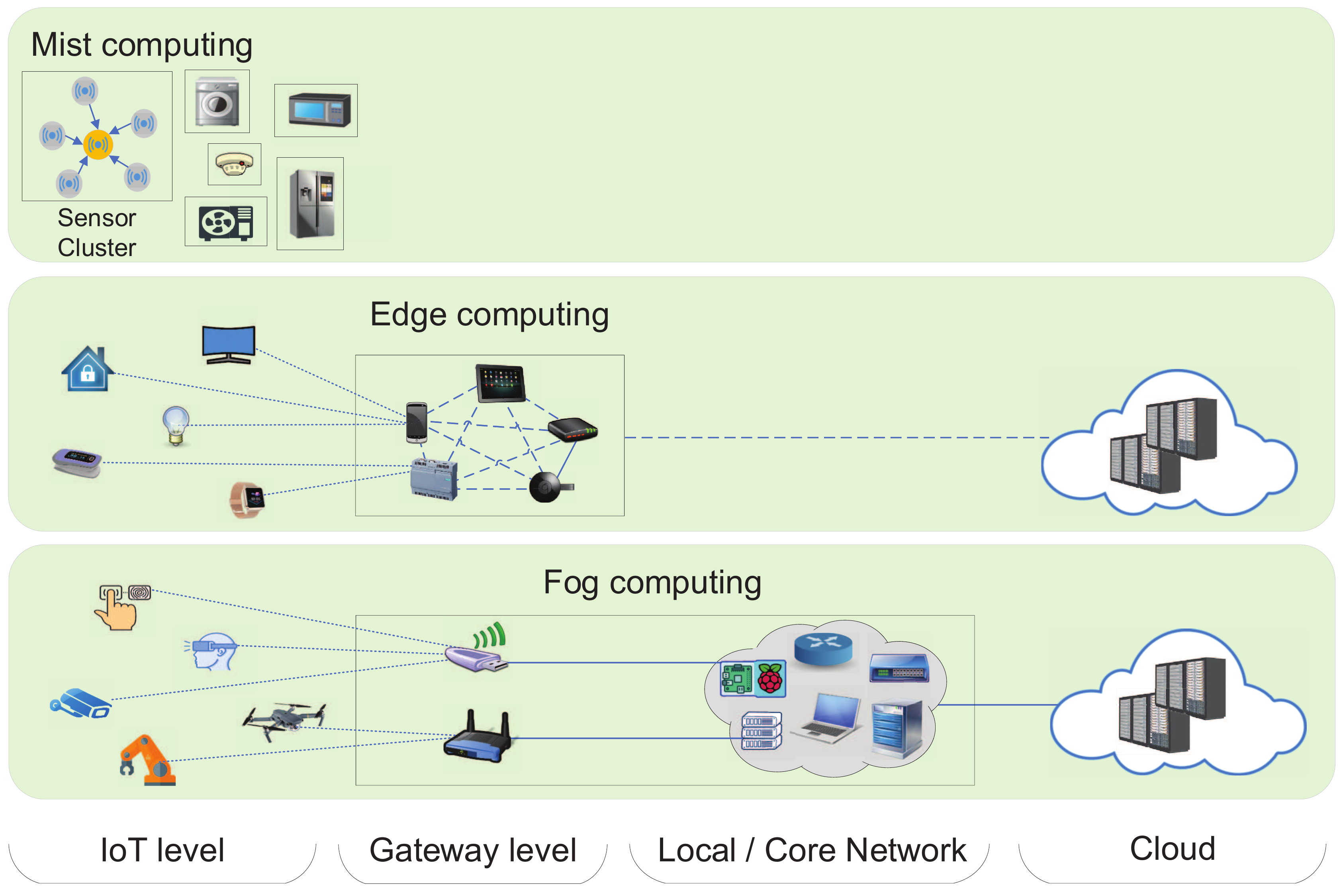}
\caption{Domain of Mist, Edge and Fog computing}
\label{Fig:domain}
\end{figure}
Likewise Fog computing, Edge and Mist computing support application execution in proximity of data sources as shown in Fig. \ref{Fig:domain}. More precisely, Mist computing enables the IoT devices to process data within themselves whereas Edge computing performs the processing operations at the gateways of IoT devices \cite{EdgeComp1} \cite{MistComp1}. For instance, smart watches can be considered as IoT devices. End users usually connect smart watches to their smart phones via Bluetooth Low Energy (LE) networking so that they can receive mobile notifications while walking or driving. Here, the smart phones act as the IoT gateways for the smart watches. At the same time, the smart watches sense blood pressure, heartbeat and oxygen saturation rate of the users. When a smart watch executes the applications to process its generated data, Mist computation occurs \cite{MistComp2}. Conversely, Edge computation happens when the smart watch forwards the data to a smart phone-based application for processing \cite{EdgeComp2}. However, compared to them, Fog computing not only harnesses the IoT gateways but also engage other computing components within the edge network such as smart routers, personal computers, Raspberry Pi devices and even micro-datacentres to process the IoT data \cite{Fog1}. 
\par Although Mist and Edge computing can solve many IoT-related issues, they have certain limitations. The computing components in Mist are not abundant in processing, networking and energy capacity. They are less capable of executing large-scale and complex applications for a longer period of time \cite{MistComp3}. On the other hand, the management of Edge nodes are very much user-centric that incorporates only reactive fault-tolerant facilities. In Edge environments, fairness is also tedious to ensure among multiple users \cite{EdgeComp3}. Fog computing overcomes these limitations of Mist and Edge by leveraging comparatively powerful resources at the user premises level and lessening the burdens of resource and application service management from the users. Moreover, Fog computing maintains a seamless communication with Cloud datacentres that eventually offers an extensive execution platform for the IoT applications \cite{Mahmud}. The notable differences between Mist, Edge and Fog computing are listed in Table \ref{computingSummary}.     
\par However, Mist, Edge and Fog are relatively new computing paradigms and their evolution processes are still ongoing. Therefore, many researchers and industries adopt different approaches to define them. For instance, there are several research works in the literature that consider Edge computing as a subset of Fog computing \cite{FogBus}. Oppositely, in other works, Edge computing is regarded as a superset embracing all paradigms where the computation is moved to the edge network, including Fog computing, Mobile Cloud computing and Mobile Edge computing \cite{EdgeComp1}. There are also some examples where Fog and Edge computing are used interchangeably \cite{31}. Moreover, in certain cases, Edge computation is regarded as a service model which is offered by different paradigms namely Dew, Mist and Fog computing. According to this concept, Dew computing happens in the IoT devices and Mist computing occurs at the IoT gateways \cite{DewComp}. Nevertheless, among these contemporary paradigms, Fog computing is considered highly feasible due to its widespread support for the IoT applications.
\begin{table}[!t]
\scriptsize
\begin{tabular}{|>{\raggedright}p{2.9 cm}>{\raggedright}p{3.2 cm}>{\raggedright}p{3.2 cm}>{\raggedright}p{3.2 cm}|}\hline
    Facts & Mist & Edge & Fog\tabularnewline \hline 
    Place of operation &  IoT devices &  Gateway devices &  Specialized networking and computing machines \tabularnewline \hdashline
    Elementary Hardware & Microcontroller &  Programmable logic controller &  Single-board computer \tabularnewline \hdashline 
    Wireless standards  &  Zigbee and Bluetooth LE &  Bluetooth and WiFi &  WiFi and LTE \tabularnewline \hdashline
    Policy manager &  Manufacturer &  Users &  Service providers \tabularnewline \hdashline 
    Application deployment & Programmed & Installed by user & Requested by user to service providers \tabularnewline \hdashline
    Resource assignment & Dedicated & Shared & Shared or virtualized \tabularnewline \hdashline 
    Application-user mapping & Single application, single user & Multiple application, single user & Multiple application, multiple user \tabularnewline \hdashline
    Resource orientation  & Standalone, homogeneous cluster & Peer to Peer, Ad-hoc & Cloud of Things \tabularnewline \hdashline
    Cloud communication & Incoherent or through mediator & Event-driven & Seamless \tabularnewline \hdashline 
    Fault tolerance techniques & Replacement & User-defined exception handling & Proactive and reactive \tabularnewline \hdashline
    Extended from & Wireless sensor network, embedded systems & Personalized computing environments & Cloud computing \tabularnewline \hline
\end{tabular}
\caption{Summary of Mist, Edge and Fog computing}\label{computingSummary}
\vspace*{-0.7cm} 
\end{table}
\subsection{Related Surveys in Fog Computing}
In the context of Fog computing, resource and application management are equally important. In fact, without efficient application service management, the capabilities of Fog resources cannot be exploited completely and vice versa. Nevertheless, in existing Fog-based literature surveys, application management is considered as a part of Fog resource management. Among these surveys, \cite{Hu}, \cite{Mouradian}, \cite{Mahmud} and \cite{Yousefpour} provide the general discussions on Fog computing. They review the researches on Fog computing from architectural perspective and highlight the key technologies and limitations of Fog computing. Moreover, they illustrate the benefits of Fog computing and clearly identify the differences between Fog and Cloud computing. Other Fog-based literature surveys including \cite{Hong}, \cite{Li} \cite{Naha} and \cite{Arani} explore the basic resource management approaches in Fog environments. They investigate various management frameworks, scheduling techniques and provisioning algorithms for Fog resources. Furthermore, they review the resource orchestration techniques in layered Fog environments and study the resource management policies in accordance with the application service requirements. There exist some other literature surveys that focus on a specific aspect of Fog resource management. For example, \cite{Osanaiye} addresses the virtual computing instances migration methods in Fog computing and \cite{Baccarelli} inspects energy-efficient Fog resource management. 
\par Moreover, \cite{Bellavista}, \cite{Nath} and \cite{Puliafito} conduct surveys to conceptualize the application service management in Fog environments. They discuss the communication, security, data and actuation management as the parts of application management. Besides, they highlight different application specific Fog architecture and give an overview to realize them for various IoT-enabled CPSs. Nevertheless, there are some other literature surveys that target particular Fog computing-based applications and their service management. For example, \cite{Aazam} study computation offloading techniques in Fog computing environments. Similary, \cite{Kraemer}, \cite{Mukherjee} and \cite{Perera} investigate the existing approaches that enable Fog computing in smart health care, advanced networking and smart city-based applications respectively. On the other hand, \cite{Roman}, \cite{Shirazi} and \cite{Zhang} are amongst those literature surveys which discuss the security aspects of Fog computing from both resource and application management perspectives. 
\par In Table \ref{Tab:related_survey}, a summary of existing Fog literature surveys and their comparative study with respect to our work is presented. As noted, the existing surveys do not explore application management in Fog environments comprehensively. More specifically, they barely discuss about application architecture, placement and maintenance in collective manner and illustrate the literature taxonomy accordingly. In this work, we address these shortcomings. We also identify the associate research gaps, demonstrate a perspective framework for distributed application management and provide future direction for the improvement of Fog computing concept. The following sections of this work present the detail review of existing application management strategies in Fog computing.   
\begin{table}[!t]
\scriptsize{
\begin{tabular}{|>{\raggedright}p{2.8 cm}| >{\raggedright} p{1.4 cm} >{\raggedright} p{1.4 cm} >{\raggedright} p{1.4 cm} >{\raggedright} p{1.4 cm} >{\raggedright} p{1.4 cm} >{\raggedright} p{1.4 cm}|}\hline
\textbf{Surveys}  & \multicolumn{6} {c|}{\textbf{Issues}}  \tabularnewline \cline{2-7}  
& Discusses application architecture & Investigates application placement techniques & Explores application maintenance operations & Provides taxonomy on application management & Conceptualizes application management framework & Relates application and resource management \tabularnewline\hline 
\cite{Aazam} & $\partial$   & \checkmark & $\partial$   & 	    & \checkmark & 	      \tabularnewline
\cite{Baccarelli} & \checkmark & 	          & 	       & $\partial$   & \checkmark & \checkmark \tabularnewline
\cite{Bellavista} & 	     & \checkmark & 	       & $\partial$   & \checkmark & \checkmark \tabularnewline
\cite{Arani} & 	      	     & \checkmark &          & $\partial$   &       & \checkmark	      \tabularnewline
\cite{Hong} & $\partial$   & \checkmark & 	       & $\partial$   & 	         & \checkmark \tabularnewline
\cite{Hu} & $\partial$   & $\partial$   & 	       & 	    & 	         & \checkmark \tabularnewline
\cite{Kraemer} & $\partial$   & \checkmark & 	       & 	    & 	         & 	      \tabularnewline
\cite{Li} & 	     & $\partial$   & $\partial$   & $\partial$   & 	         & 	      \tabularnewline
\cite{Mahmud} & \checkmark & \checkmark & $\partial$   & $\partial$   & 	         & 	      \tabularnewline
\cite{Mouradian} & 	     & $\partial$   & 	       & 	    & \checkmark & \checkmark \tabularnewline
\cite{Mukherjee} & $\partial$   & \checkmark & \checkmark & 	    & \checkmark & 	      \tabularnewline
\cite{Naha} & 	     & $\partial$   & $\partial$   & \checkmark & \checkmark & \checkmark \tabularnewline
\cite{Nath} & 	     & $\partial$   & \checkmark & $\partial$   & \checkmark & 	      \tabularnewline
\cite{Osanaiye} & 	     & 	          & $\partial$   & 	    & \checkmark & \checkmark \tabularnewline
\cite{Puliafito} & $\partial$   & \checkmark & 	       & 	    & 	         & \checkmark \tabularnewline
\cite{Perera} & \checkmark & \checkmark & 	       & 	    & \checkmark & 	      \tabularnewline
\cite{Roman} & $\partial$   & 	          & \checkmark & 	    & 	         & \checkmark \tabularnewline
\cite{Shirazi} & 	     & 	          & \checkmark & 	    & \checkmark & \checkmark \tabularnewline
\cite{Yousefpour} & 	     & \checkmark & \checkmark & $\partial$   & 	         & 	      \tabularnewline
\cite{Zhang} & $\partial$   & 	          & \checkmark & 	    & \checkmark & \checkmark \tabularnewline
This survey & \checkmark & \checkmark & \checkmark & \checkmark & \checkmark & \checkmark \tabularnewline \hline
\end{tabular}} 
\begin{tablenotes}
     \item[1] $\checkmark$ denotes broad discussion on the respective issue.
     \item[2] $\partial$ denotes partial discussion on the respective issue.
   \end{tablenotes}
\caption{Summary of literature surveys in Fog computing}\label{Tab:related_survey}
\vspace*{-0.7cm} 
\end{table}
\section{Application Architecture} \label{Sec_Design}
\begin{figure}[!t]
\centering 
\captionsetup{justification=centering,margin=2cm}
\includegraphics[width=130mm, height= 45mm]{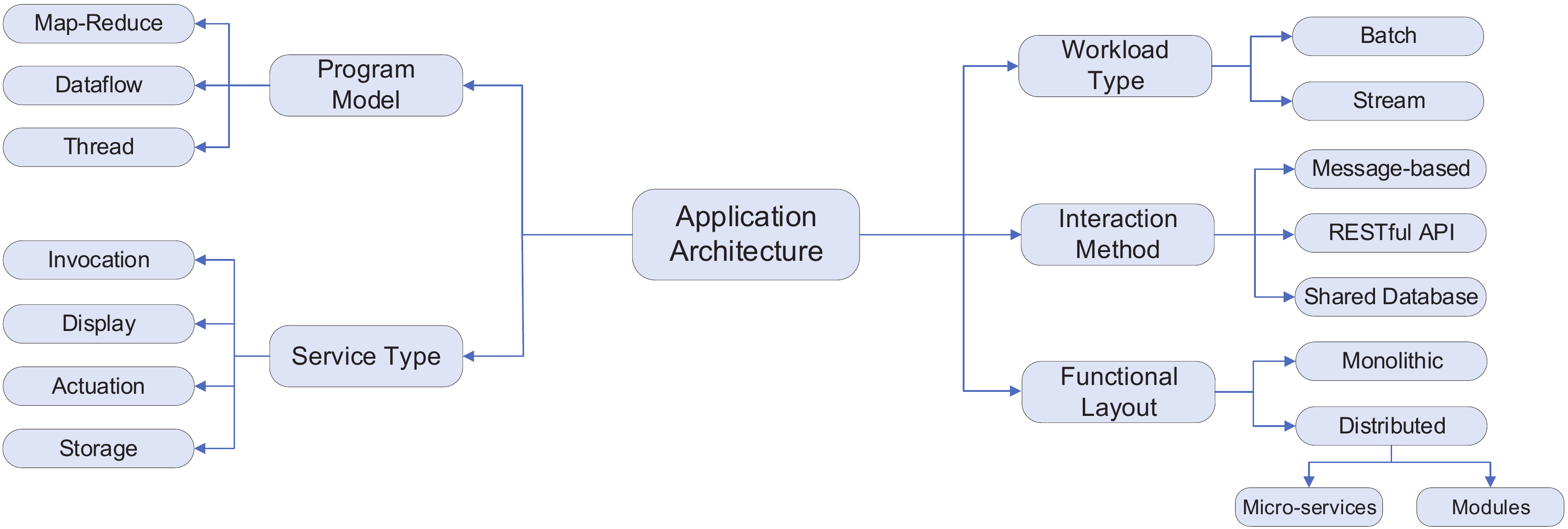}
\caption{Taxonomy on application architecture}
\label{Fig:design}
\end{figure}

The complexities of executing IoT applications in distributed, heterogeneous and resource constrained Fog nodes can be addressed if the architecture of applications is defined as per the specifications of corresponding Fog environment. An elastic architecture also helps interoperability between different versions of an applications. Moreover, the elements of application architecture such as programming model and workload type are used to determine the placement strategy and resource consumptions of the applications. The service type of an application denotes the scope of its external exposure that assists in application maintenance. Fig. \ref{Fig:design} provides a taxonomy on application architecture highlighting the main elements. These elements are described below.   
%
%
\subsection{Functional Layout}
An application performs different types of operations. For example, an image processing application reduces noises from an image, converts colors, extracts features and compares the results with predefined thresholds. The functional layout of an application defines the arrangement of these operations and assists in realizing the possible distribution of the application in constrained Fog environments. The functional layout of applications can be classified into two types;
\subsubsection{Monolithic:}
In monolithic applications, all computational operations are encapsulated in a single program. These applications function independently to each other. Within such applications, developer specific parallelism techniques are applied so that they can run over multiple processing cores of the host Fog node. In literature, \cite{1}, \cite{9}, \cite{36}, \cite{39} and \cite{102} discuss about the monolithic applications in Fog.      
\subsubsection{Distributed:}
The computational operations of a distributed application are organized in separate programs. Compared to monolithic applications, distributed applications are easier to expand. The programs of a distributed applications can be executed in a single Fog node or can be operated by several Fog nodes in collaborative manner. Based on the dependency of computational operations, distributed applications are classified into two categories.
\par $\bullet$ \textit{Module-based:} In module-based distributed applications, the application programs are tightly coupled and dependent to each other. They are devotedly provisioned for serving data of a particular source. \cite{2}, \cite{11}, \cite{15}, \cite{16} and \cite{76} discuss about module-based distributed applications.  
\par $\bullet$ \textit{Micro-services:} Through micro-service-based implementation, the computational operations of an application are shared among different CPSs to process their data simultaneously. Unlike application modules, micro-services are loosely coupled and function independently. Due to explicit isolation, a micro-service of an application can be easily attached to other applications as per the requirements. \cite{46}, \cite{51}, \cite{65}, \cite{96} and \cite{114} highlight the micro-service-based implementation of IoT applications.  
%
%
\subsection{Program Model}
Program Model defines the execution order of computing operations in an application and guides to provision resources for application as per their dimension and predicted life cycle. Three different types of program models have been widely adopted while developing the applications in Fog. 
\subsubsection{Thread:}
To ensure simultaneous execution of independent computational operations within an application, thread program model is used. It is one of the primitive program models that helps to achieve concurrency in resource constrained Fog nodes. \cite{10}, \cite{31}, \cite{47}, \cite{55}, \cite{53} and \cite{88} follow thread model to develop applications for Fog computing environments. The advanced versions of thread model such as map-reduce and dataflow are also used predominantly in Fog computing.    
\subsubsection{Map-Reduce:}
Through this model, the large-volume inputs for an application is divided into multiple chunks so that its all operations can run in parallel over the given inputs. Later, the processing outcomes of each chunk are combined to generate the overall output of the application. Such program model for Fog-based applications is discussed in \cite{103}, \cite{MR_1}, \cite{MR_2}, \cite{MR_3}, \cite{MR_4} and \cite{MR_5}.   
\subsubsection{Dataflow:} 
In dataflow program model, the output of a computational operation is fed to another operation as input and this process continues for the subsequent operations. This program model binds all computational operations of an application through a Directed Acyclic Graph (DAG). The size of input data handled through this model is not usually as large as that of the map-reduce model. In literature \cite{68}, \cite{74}, \cite{75}, \cite{66}, \cite{85} and \cite{38} consider dataflow program model for the applications.
%
%
\subsection{Service Type}
The service type of an application refers to its outcome that is delivered to the corresponding and requesting CPSs. The size of output of an application depends on its service types which helps to model the data propagation delay. Based on the physical environment, application service outcomes can vary. The services of different IoT applications can be classified into four types.
\subsubsection{Invocation:}
An IoT application can invoke the execution of another application as its service. For example, the IoT application monitoring forest fire can initiate a robotic application to meet the emergency requirements. Usually, in this type of services, an initiation command along with necessary arguments are forwarded from the source application to the requested application. \cite{11}, \cite{19} and \cite{65} discuss invocation as service type of the applications.    
\subsubsection{Display:}
There exist several IoT applications such as virtual reality gaming and smart surveillance that visualize the service outcomes to the users. The quality of such application services explicitly depends on the networking condition between the end users and the associated Fog nodes. In literature, \cite{12}, \cite{21}, \cite{18}, \cite{101} and  \cite{104} highlight display-based services of applications in Fog environments.    
\subsubsection{Actuation:}
After processing incoming data, several IoT applications trigger actuators to initiate the required physical action. For example, the remote patient monitoring system can actuate the Oxygen supply engine during emergency situations. \cite{1}, \cite{34}, \cite{48} and \cite{103} consider actuation as a service type for the applications. 
\subsubsection{Storage:}
For long-term data collection or crowd sourcing, IoT applications are often used. These applications aggregate these data and store in Cloud or Fog nodes for future analysis by other applications. \cite{2}, \cite{33} and \cite{45} mention storage as an application service in Fog computing. 
%
\subsection{Interaction Method}
While processing data in a collaborative manner, the computing operations and applications require interactions with each other to share and store the intermediate outcomes for further usages. However, such interactions become very crucial when it operate across multiple Fog nodes. In the following subsections, different interaction methods for Fog-based applications are discussed.     
\subsubsection{Shared Database:}
It is one of the primitive methods of sharing data. In this method, data is stored in a particular location and all the applications and computing operations requiring the data have direct access to it. This method also supports multi-level data distribution from local and global perspectives for large-scale systems. Shared database is discussed in \cite{12}, \cite{45}, \cite{67} and \cite{85} for Fog-based applications.
\subsubsection{Message-based:}
In this method, the host Fog nodes of computing operations or applications exchange lightweight messages to notify the current state of data processing. In most of the cases, this message transmission is supervised by a dedicated entity and follows the publish and subscribe protocol for machine to machine communication. Unlike shared database-driven interaction, this method is often used for small-scale systems. In literature, \cite{57}, \cite{63}, \cite{82} and \cite{75} discuss message-based interactions in Fog.   
\subsubsection{Representational State Transfer:}
It establishes a web service-based communication between the host Fog nodes using http protocol. Representational state transfer allows data exchange through several stateless commands such as get and post, and often follows the push and pull approach during device level interactions. \cite{60}, \cite{79}, \cite{104}, \cite{FogBus} and \cite{Latency} consider representational state transfer in their works. Due to the speed and ease of scalability, this method is being widely used by the IoT applications compared to shared database and message-based interactions.
%
%
%
\subsection{Workload Type}
Workload denotes the characteristics of input processed by an application. The knowledge of workload is very important to select the host Fog node having the appropriate configurations. There are two types of workload for IoT applications
\subsubsection{Batch:}
The bundle of non-interactive inputs for an application is often regarded as Batch workload. Once accumulated from multiple sources, the batch workload is submitted to the application for processing. The dispatch order of the inputs in such workload can be shuffled as per the availability of resources to ensure the desired performance of the application. In literature, \cite{66}, \cite{103}, \cite{91}, \cite{94}, \cite{88} and  \cite{114} discuss about batch workload for the Fog-based applications.    
\subsubsection{Stream:}
This type of workload is generated by different sources in periodic manner. Therefore, while developing real-time IoT solutions, the stream workload is preferred more than the batch workload. The specification and processing requirements of stream workload can change with the course of time based on the sensing frequency of associated IoT devices. \cite{67}, \cite{70}, \cite{71}, \cite{89} and \cite{38} discuss about stream workload in Fog.   
\subsection{Research Gaps in Application Architecture} 
Table \ref{Tab:related_appDesign} summarizes the existing concepts related to application architecture in Fog computing. Although there are a notable number of works, some issues related to this aspect of application management are yet to be investigated. They are discussed below:
%
%
\begin{landscape}
{\centering
{\scriptsize
\begin{longtable}{|>{\raggedright}p{3 cm}| >{\raggedright} p{1 cm} >{\raggedright} p{1 cm} >{\raggedright} p{1 cm} >{\raggedright} p{1 cm} >{\raggedright} p{1 cm}||
>{\raggedright}p{3 cm}| >{\raggedright} p{1 cm} >{\raggedright} p{1 cm} >{\raggedright} p{1 cm} >{\raggedright} p{1 cm} >{\raggedright} p{1 cm}|}\hline
Works  & \multicolumn{5} {c||}{Application Architecture} & Works  & \multicolumn{5} {c|}{Application Architecture} \tabularnewline \cline{2-6}\cline{8-12}  
& Program Model & Service Type & Workload Type & Interaction Method & Functional Layout & & Program Model & Service Type & Workload Type & Interaction Method & Functional Layout \tabularnewline\hline 
\cite{1} &   &  Actuation  & Batch  &   & Monolithic & \cite{58} & Dataflow  &   &  Stream &   &  Module\tabularnewline\hdashline 
\cite{2} & Dataflow & Storage & Batch  &   &  Module & \cite{60} &  Dataflow  &   &   &  REST &  Module \tabularnewline\hdashline
\cite{47} &  Dataflow, Thread  &   & Stream &   &   & \cite{61} &   &   & Batch  &   & Monolithic  \tabularnewline\hdashline 
\cite{10} & Thread &   & Stream &   &  & \cite{65} &   & Invocation &   &  &  $\mu$-service \tabularnewline\hdashline 
\cite{11} &   & Invocation  &   &   &  Module & \cite{66} &  Dataflow  &   & Batch  &   &  Module \tabularnewline\hdashline
\cite{12} &   &  Display  & Stream & Shared Database & Monolithic & \cite{68} &  Dataflow  &  Actuation  &   &   &  \tabularnewline\hdashline
\cite{14} &   &   &  Stream &   & Monolithic & \cite{70} &   &   &  Stream &   & Monolithic  \tabularnewline\hdashline 
\cite{21} &  Dataflow  &  Display &   &  & Module & \cite{67} & Map-Reduce  &  Display  &  Stream & Shared Database  & Monolithic \tabularnewline\hdashline 
\cite{24} &   &  & Batch &  & Monolithic & \cite{71} &  Thread &   & Stream &   &   \tabularnewline\hdashline   
\cite{31} &  Thread &   & Batch  &   & Monolithic  & \cite{74} &  Dataflow  &  Actuation  &   &   & Module \tabularnewline\hdashline 
\cite{32} &   &  Display  & Batch  &   &   & \cite{75} &  Dataflow &   &  & Message  &  Module \tabularnewline\hdashline   
\cite{33} & Dataflow & Storage &  &  & Module & \cite{76} &   & Display  &   &   & Module \tabularnewline\hdashline    
\cite{34} &   &  Actuation  & Batch  &   &  Module & \cite{79} &   &   &   &  REST  &  $\mu$-service  \tabularnewline\hdashline 
\cite{35} &   &   & Stream &   & Monolithic  & \cite{80} &  Dataflow  &   &   &   &  Module \tabularnewline\hdashline 
\cite{15} &  Dataflow  & Invocation, Display &  Stream &   & Module & \cite{82} &  Dataflow  &   & Batch  & Message  &  $\mu$-service \tabularnewline\hdashline     
\cite{16} &  Dataflow &  & & & Module & \cite{83} &   &   &  Stream &   & Monolithic  \tabularnewline\hdashline 
\cite{18} &   & Display & Batch &  & & \cite{85} &  Dataflow &   &   & Shared Database &  Module \tabularnewline\hdashline  
\cite{19} &   & Invocation  & Batch &  & Monolithic & \cite{88} &  Thread &   & Batch  &   &  \tabularnewline\hdashline 
\cite{28} &  Dataflow  &  Display  &   &  & Module & \cite{89} &   &   &  Stream &   &  $\mu$-service \tabularnewline\hdashline 
\cite{36} &   &   & Batch  &   & Monolithic & \cite{90} &   &   &  Stream &   & Monolithic \tabularnewline\hdashline 
\cite{108} &  Dataflow  &   &   &   &  Module & \cite{63} &  &   &  Stream & Message  &   \tabularnewline\hdashline 
\cite{37} &  Dataflow  &  Display  &  Stream &   &  Module & \cite{91} &   &   & Batch  &   & Monolithic \tabularnewline\hdashline 
\cite{38} &  Dataflow  &   &  Stream &   &  $\mu$-service & \cite{94} &   &   & Batch  &   & Monolithic \tabularnewline \hdashline
\cite{39} &   &   &  Stream &    & Monolithic  & \cite{96} &  Dataflow  &  &   &  REST  &  $\mu$-service \tabularnewline\hdashline 
\cite{44} & Dataflow &   &  Stream &   &  Module & \cite{97} & Dataflow  &   & Batch  &   &   \tabularnewline\hdashline 
\cite{45} &   & Storage &  Stream & Shared Database & Monolithic  & \cite{100} &   &   & Batch  &   & Monolithic  \tabularnewline\hdashline 
\cite{46} &  Dataflow  &   &   &   &  $\mu$-service & \cite{101} &   &  Display  & Batch  &  & Monolithic \tabularnewline\hdashline 
\cite{48} &  Dataflow  &  Actuation  &   &   &  Module & \cite{102} &  &   &  Stream &   & Monolithic \tabularnewline\hdashline 
\cite{51} &  Dataflow  &   &   &   &  $\mu$-service  & \cite{103} & Map-Reduce  &  Actuation  & Batch  &   &  \tabularnewline\hdashline     
\cite{53} & Thread &  Display & Batch  &   & Monolithic & \cite{104} &  Dataflow  &  Display  & Stream &   &  $\mu$-service \tabularnewline\hdashline 
\cite{55} &  Thread &   & Batch  &   &   & \cite{112} &   &   & Batch  &   & Monolithic \tabularnewline\hdashline  
\cite{56} &   &   &  Stream &   & Monolithic & \cite{114} &   &   & Batch  &   &  $\mu$-service \tabularnewline
\hline 
\addlinespace[1ex]
\caption{Summary of existing concepts on application architecture}\label{Tab:related_appDesign}
\end{longtable}}}
\end{landscape}
%
%
%
\par 1. The execution of one application having particular programming model can trigger another application with different programming model. In such cases, the dynamic reconfiguration of Fog nodes is required. However, in existing works, only the static configuration of Fog nodes have been discussed \cite{cloudLab2} \cite{13}.
\par 2. The varying service type of applications can affect the networking capabilities of the host Fog nodes and degrade the service time the applications. Nevertheless, the existing approaches barely consider multiple service types of an applications simultaneously while determining a suitable placement options for them \cite{Edge}.
\par 3. There are some research works denoting that the higher sensing frequency of IoT devices is required for better accuracy \cite{FogBus}. However, they have not considered that the high streaming rate of data creates immense processing burden on the Fog nodes.
\par 4. Although, monolithic applications alleviate the constraints of inter-nodal communication delay, they are less modular. Conversely, the distributed application offer scalability but their service often gets affected by the limitations of underlying network. Although dynamic modularization of applications as per the context of Fog network is required, current researches only focus on the fixed functional layouts \cite{21}.
\section{Application Placement} \label{Sec_Placement}
\begin{figure}[!t]
\centering 
\includegraphics[width=140mm, height= 100mm]{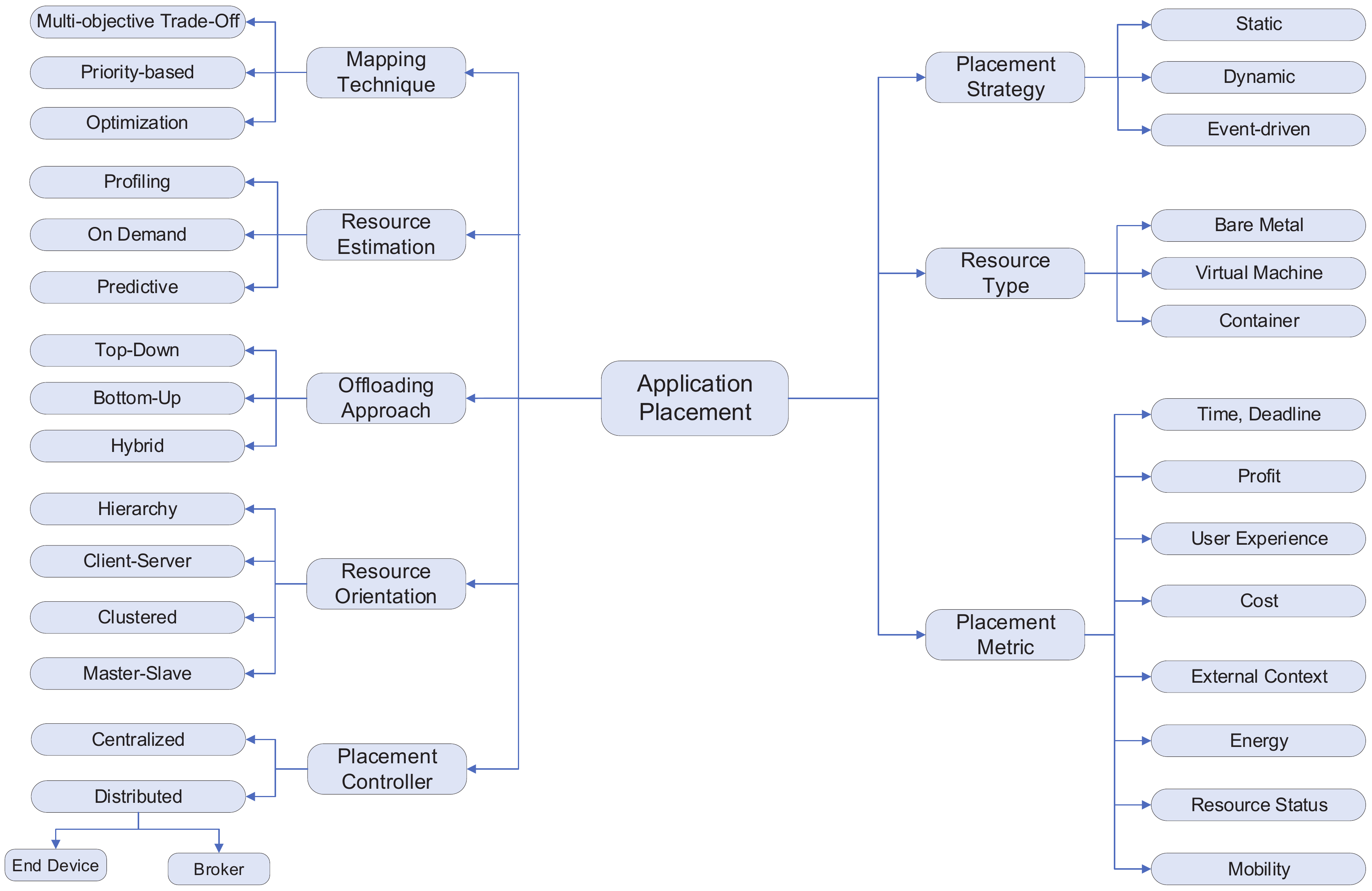}
\caption{Taxonomy on application placement}
\label{Fig:placement}
\end{figure}

The task distribution problem in Fog computing can be solved to a great extent if the applications are placed considering the future processing commitments of the Fog nodes. Additionally, the opportunistic placement of the applications can be a potential factor to standardize the Fog and Cloud integration. Moreover, while placing the applications, the resource orientation and their status are studied extensively. Such studies can play a vital role to update the application architecture dynamically and ensure proactive application maintenance. Fig. \ref{Fig:placement} depicts a taxonomy of various elements relevant to the application placement. Their descriptions are given below.
%
%
\subsection{Resource Estimation}
Resource estimation helps to determine whether an application is compute-intensive, I/O intensive or disk intensive. This information is crucial to make placement decisions with respect to the capacity constraints of Fog nodes. There are three types of resource estimation techniques in Fog.
\subsubsection{Profiling:}
When a limited number of Fog nodes reside in a Fog environment and the specifications of requested applications remain static, the profiling technique is predominantly used to estimate the resources for an application. In this technique, an application is executed separately on each Fog node and associated performance parameters such as processing time, propagation time and energy consumption are monitored. Based on the accumulated information, the suitable resources for the subsequent executions of the application are selected. In \cite{18}, \cite{48}, \cite{114} and \cite{115} application profiling is discussed.
\subsubsection{Predictive:}
In this technique, based on the past execution patterns, the appropriate resources for an application are determined. This technique is highly applicable when a Fog environment supports dynamic provisioning of its component Fog nodes and the specifications of requested application vary. Compared to the profiling of applications, this technique is highly scalable, however, its results can be less precise. On the other hand, profiling depends on the physical deployment whereas prediction relies on the mathematical implications. \cite{23}, \cite{27}, \cite{33} and \cite{106} discuss predictive resource estimation.
\subsubsection{On Demand:}
In this technique, resources are estimated based on the expectations of users and their instant demand. This technique differs from profiling or predictive techniques where the resource estimation depends on the application characteristics. \cite{14}, \cite{22}, \cite{49} and \cite{FogBus} consider on-demand resource estimation. 
%
%
\subsection{Offloading Approach}
An offloading approach helps to manage the applications and their associated data as per the up-link and down-link network overhead of the host Fog nodes. In literature, three types of offloading approaches have been found for the Fog-based applications. 
\subsubsection{Bottom-Up:}
In this type of offloading, the requests regarding application services and relevant data are directly forwarded to the Fog nodes from the IoT devices or end users. This approach for Fog-based applications are highlighted in \cite{56}, \cite{59}, \cite{60} and  \cite{95}. 
\subsubsection{Top-Down:}
Unlike bottom-up offloading, the top-down approach pushes the workload and programs of an application from Cloud to Fog nodes as per the requests of the end users. This technique is often applied when Fog nodes are used for content distribution. \cite{4}, \cite{48}, \cite{52} and \cite{113} consider this approach for Fog.
\subsubsection{Hybrid:}
Apart from the aforementioned offloading approaches, Fog nodes can share application programs among themselves. During such interactions, a hybrid pattern of top-down and bottom-up offloading is followed. This approach is considered in \cite{63}, \cite{73}, \cite{94} and  \cite{114} for Fog environments. 
%
%
\subsection{Resource Orientation}
When different applications work collaboratively and deployed in a distributed manner across multiple Fog nodes, the inter-nodal communication delay becomes a very important factor to meet their QoS. This delay largely depends on how the computing resources are arranged in Fog environments. There are four types of resource orientation in Fog. 
\subsubsection{Hierarchy:}
In this orientation, Fog nodes are arranged in different hierarchical levels between the communication path of IoT devices and Cloud datacentres. In the lower level, the number Fog nodes is higher compared to that of the higher levels. Conversely, as the level number goes higher, the inter-nodal communication delay between the devices increases. This orientation assists in vertical scaling of the resources. In \cite{1}, \cite{2}, \cite{5}, \cite{12} and \cite{115}, this type of orientation of Fog nodes is highlighted.
\subsubsection{Clustered:}
In clustered resource orientation, all Fog nodes are directly or logically connected to each other and share the information among themselves with high throughput communication channels. In most of the cases, the external communication, resource management and resource discovery operations within a Fog cluster are supervised by a dedicated Fog node. The clustered orientation provides more scope for horizontal scaling than the hierarchical orientation. This resource orientation is discussed in \cite{9}, \cite{24}, \cite{32}, \cite{cloudLab2}, \cite{46} and \cite{55}. 
\subsubsection{Client-Server:}
Client-Server-based resource orientation enables a set of Fog nodes to work as the servers while lets others to act as the clients. The client Fog nodes request the server Fog nodes to process their forwarded data. This orientation is often regarded as a combination of clustered and hierarchical orientation. \cite{3}, \cite{7}, \cite{8}, \cite{13}, \cite{14}, and \cite{31} discuss the client-server orientation of Fog resources. 
\subsubsection{Master-Slave:}
In this orientation, a master Fog node distribute the data processing responsibilities to other slave Fog nodes and explicitly manages their operations. After receiving the service outcomes from the slave nodes, the master node accumulates them and forward the final results to the destination as per the service type. This orientation is more efficient in distributing the computing responsibilities than the client-server orientation. In \cite{20}, \cite{22}, \cite{39}, \cite{79}, \cite{67} and  \cite{FogBus}, the master-slave resource orientation is considered.
%
%
\subsection{Placement Controller}
Placement controller defines the logical location of an entity that manages the overall application management operations in Fog computing. Furthermore, it assists in estimating the waiting time from requesting to placing an application in Fog environment which consequently drives the overall performance of the IoT-enabled CPSs. There are two types of placement controller in Fog.  
\subsubsection{Centralized:}
This type of placement controller locates in a commonly accessible place by the Fog nodes and poses a global view of the Fog environment. Generally, they are hosted in Cloud datacentres and supervise the Fog nodes residing at the edge network. \cite{33}, \cite{48}, \cite{52}, \cite{79}, \cite{64} and\cite{91} discuss about the centralized placement controllers of Fog environments. 
\subsubsection{Distributed:}
Unlike the centralized controller, the distributed placement controllers manage the Fog nodes based on a local view of the Fog environments. They are classified into two categories. 
\par $\bullet$ \textit{End Device: }In this type of distributed placement controller, the IoT devices and the Fog nodes not only perform their predefined responsibilities such as data sensing and data processing, but also take the application management decisions for other Fog nodes. \cite{54}, \cite{55}, \cite{56}, \cite{57}, \cite{63} and \cite{78} consider end devices as distributed placement controller for Fog environments.
\par $\bullet$ {Broker: }In contrast to end devices, this type of controllers are considered dedicated for application management operations in Fog environments. The brokers reside in proximity of the Fog nodes and interact with the external entities on behalf of the Fog nodes and vice-versa. \cite{29}, \cite{30}, \cite{39}, \cite{41}, \cite{82} and \cite{84} discuss broker based placement controllers in Fog environments.
%
%
\subsection{Mapping Technique}
Based on different parameters, an application placement policy provides the mapping of the applications with respect to the Fog nodes and their virtualized instances. The complexity of the adopted mapping technique defines the runtime of the policy which consequently denotes its responsiveness. Three different types of mapping techniques are commonly used in Fog computing.
\subsubsection{Priority-based:}
This technique prioritizes an application placement on particular Fog node or virtualized instances. Generally, the heuristic approaches such as best fit and first fit are commonly used for prioritization of the applications. In \cite{7}, \cite{11}, \cite{13}, \cite{53}, \cite{79} and \cite{80}, prioritization is highlighted as the mapping technique.
\subsubsection{Optimization:}
This technique either maximizes or minimizes one particular objective function while placing the applications in Fog environments. Although optimization provides the best mathematical solution of a problem, this technique takes more time to operate than prioritization. Different types of optimization approaches such as linear, non-linear and constrained are widely studied in Fog computing. \cite{16}, \cite{18}, \cite{21}, \cite{23}, \cite{34} and \cite{115} discuss this mapping technique for Fog. 
\subsubsection{Multi-objective Trade-off:}
Unlike optimization, the multi-objective trade-off can maximize or minimize two or more objectives such as time-energy, time-cost and cost-QoE simultaneously while placing the applications. Different meta-heuristic approaches including particle swarm, evolutionary algorithms, game theory and multi-objective optimization are used for making trade-off among various application placement metrics. \cite{2}, \cite{20}, \cite{24}, \cite{27} and \cite{64} consider trade-off for placing the applications. 
%
%
\subsection{Placement Strategy}
The iterative execution of applications in Fog environments depends on the arrival of their inputs, or data processing life cycle. The placement algorithms need to consider these issues so that they can detect suitable hosts for different application. Placement strategy helps to define how frequently the placement algorithms are required to be executed for subsequent execution of an application. There are three types of placement strategies for Fog computing. 
\subsubsection{Static:}
In this strategy, placement algorithm is executed once for each application and at the host, the application is kept running. Inputs of an application are directly sent to its host from the IoT devices as the processing destination remain same for all of them. \cite{3}, \cite{5}, \cite{8}, \cite{34}, \cite{35} and \cite{54} discuss about the static application placement strategy in Fog computing environments. 
\subsubsection{Dynamic:}
In Fog, an application can have multiple replicas running or the application can terminate by processing only a single input. For both cases, the placement algorithm requires to be executed for each arrival of the inputs to determine where to schedule them or where to execute the application next. Such dynamic of placement strategy is highlighted in \cite{1}, \cite{4}, \cite{46}, \cite{50} and \cite{107}.
\subsubsection{Event-driven:}
An application often requires to be relocated from one host to another. This relocation can be occurred because of the mobility of the sources and destinations, preemption, Fog node consolidation or service migration. In such cases, after initial placement, further scheduling of applications is conducted occasionally based on the occurrence of the event. Event driven strategy for Fog-based application placement is considered in \cite{19}, \cite{21}, \cite{53}, \cite{77} and  \cite{114}.
%
%
\subsection{Resource Type}
Fog nodes contain necessary resources such as processing cores, memory and bandwidth to assist the execution of applications. They can be virtualized to support multi-tenancy on the physical resources. Resource type denotes the internal features of the host of the applications that helps in validating the scope of dynamic allocation of resources during application runtime. Three different types of resources are discussed in Fog computing.
\subsubsection{Bare Metal:}
In this type of resources applications are directly placed to the Fog nodes without any virtualization. Applications access the physical hardware of Fog nodes through the host operating system. Such type of resources can support multi-tenancy without explicit isolation of the application execution unit. \cite{18}, \cite{19}, \cite{21}, \cite{22}, \cite{31} and \cite{70} highlight bare metal as resource type of Fog nodes. 
\subsubsection{Virtual Machine:}
In contrast to bare metal resources, virtual machines exploits hardware level virtualization so that multiple operating systems can run independently on a single Fog node. They run on top of an abstraction layer named hypervisor that enables the sharing of hardware among different virtual machines. \cite{14}, \cite{26}, \cite{29}, \cite{30} and \cite{64} consider application placement in virtual machines.   
\subsubsection{Container:}
This type of virtualized resources is lightweight compared to virtual machines and offers operating-level virtualization. In opposition to bare metals, containers isolate processes with required application packages and they are highly portable across multiple Fog nodes. Containers are used in \cite{20}, \cite{49}, \cite{57}, \cite{79}, \cite{cloudLab1} and \cite{96} for application placement in Fog environments.
\subsection{Placement Metric}
The main intention of placing applications in Fog can vary according to the requirements of users, service providers and physical environments. Placement metric refers to the parameters that set the objectives of application placement in Fog environments. A wide variety of placement metrics are noted in Fog computing. They are described below. 
\subsubsection{Time and Deadline:}
This placement metric signifies the aim of minimizing application service delivery time and meeting the specified deadline. While setting this metric, the computation time, data propagation time and node deployment time are also considered. \cite{7}, \cite{18}, \cite{21}, \cite{25}, \cite{40} and \cite{50} discuss time as the placement metric in Fog.
%
%
\subsubsection{Profit:}
Service providers get benefited when the applications are deployed in Fog with a view to maximizing their profit and revenue. The intention of making profit often leads the providers to offer application execution in Fog as a utility. \cite{9}, \cite{26}, \cite{91}, \cite{Profit} and \cite{113} consider profit as the placement metric. 
\subsubsection{User Experience:}
Service requirements of users and their affordability level can change with the course of time. If these issues are not met during the application placement, user experience can degrade. This event can also resist the users to execute applications through Fog computing in future. \cite{13}, \cite{22}, \cite{29}, \cite{QoE} and \cite{59} highlight user experience as the placement metric.
\subsubsection{Cost:}
There are different monetary costs such as infrastructure deployment cost, operational cost and instance rental cost are associated with Fog computing. Cost as placement metric refers to its minimization during the application placement in Fog. This metric is considered in \cite{12}, \cite{15}, \cite{24}, \cite{32}, \cite{39} and \cite{115} to place the applications in Fog. 
\subsubsection{External Context:}
There exist several external parameters including the relinquish rate and the activity of users, reliability of Fog nodes, the popularity of application services, data size and the sensing frequency of IoT devices that drive the decision of application placement in Fog. \cite{8}, \cite{10}, \cite{11}, \cite{27}, \cite{33} and \cite{114} consider such external contexts while placing applications in Fog.

\subsubsection{Energy:}
Fog nodes can utilize both renewable and non-renewable energy to operate. However, the energy consumptions of Fog nodes are subjected to the environmental and supply-demand related issues. \cite{2}, \cite{3}, \cite{6}, \cite{16} and \cite{41} highlight energy as a placement metric for the application. 
\subsubsection{Resource Status:}
Fog nodes are widely heterogeneous in terms of their processing power, networking interfaces, storage capacity and operational platform. Assessment of these status parameters is very important to efficiently manage the applications over them. \cite{1}, \cite{5}, \cite{6}, \cite{13}, \cite{14} and \cite{34} give higher preferences to resource status while placing the applications in Fog environments. 
\subsubsection{Mobility:}
In the context of Fog computing, both the IoT devices and the Fog nodes can move from one location to another very frequently. This feature of Fog computing can affect the service delivery and occur migration of application execution among the Fog nodes. Taking cognizance of these issues, mobility is considered in \cite{29}, \cite{30}, \cite{60}, \cite{72}, \cite{80} and \cite{114} for Fog computing.  
%
%
\subsection{Research Gaps in Application Placement} 
Table \ref{Tab:related_appPlacement} summarizes the existing application placement techniques in Fog computing. Although an extensive amount of research has been conducted on this aspect of application management, there are still some gaps. They are discussed below:
\par 1. For remote areas, many researches suggest to use renewable power sources to run the Fog nodes as the grid-based energy is costly to facilitate \cite{70}. However, very few of them consider that the availability of renewable energy is subjected to uncertainty and environmental context and take the required measures to solve the problem.
\par 2. The distribution of application placement tasks across multiple entities can reduce the management overhead. However, the existing works have not considered the elevation in decision-making time that can occur due to such distribution \cite{94} \cite{cloudLab1}.
%
%
%
\begin{landscape}
{\centering
{\scriptsize
\begin{longtable}{|>{\raggedright}p{3.5 cm}| >{\raggedright} p{1.4 cm} >{\raggedright} p{1.3 cm} >{\raggedright} p{1.3 cm} >{\raggedright} p{2 cm} >{\raggedright} p{2 cm}>{\raggedright}p{1.5 cm} >{\raggedright} p{1.5 cm} >{\raggedright} p{2.65 cm}|}\hline
Works  & \multicolumn{8} {c|}{Application Placement} \tabularnewline \cline{2-9}
& Mapping \newline Technique & Resource Estimation & Offloading Approach & Resource Orientation & Placement Controller & Placement\newline Strategy & Resource\newline Type & Placement \newline Metric \tabularnewline\hline 
\cite{1} &   &   &   & Hierarchy  & Centralized  &  Dynamic  &  VM  & Time, Resource\tabularnewline\hdashline 
\cite{2} & Trade-off  &   &  Bottom-Up  & Hierarchy  &  Broker &   & Bare Metal  & Time, Energy\tabularnewline\hdashline 
\cite{130} &  Priority  &   &  Hybrid & Hierarchy  &  End Device & Static  & Bare Metal  & Time, Resource \tabularnewline\hdashline 
\cite{3} &  Optimization &   &  Bottom-Up  &  Client-Server  &  Broker & Static  & Bare Metal  & Time, Energy\tabularnewline\hdashline 
\cite{4} & Optimization &   & Top-Down  &   &   &  Dynamic  &   &  QoE \tabularnewline\hdashline 
\cite{5} &  Optimization &   &   & Hierarchy  & End Device  & Static  & Bare Metal  & Time, Resource\tabularnewline\hdashline 
\cite{47} &  Priority  &   & Hybrid &  Cluster  &  End Device  &  Dynamic  & Bare Metal  & Time, Resource \tabularnewline\hdashline 
\cite{6} &   &   &  Bottom-Up  & Hierarchy  &  End Device  &  Dynamic  &  VM  &  Energy, Resource \tabularnewline\hdashline 
\cite{7} &  Priority  &  &   &  Client-Server  & Centralized  &  Dynamic  & VM & Time \tabularnewline\hdashline 
\cite{8} &   &   & Top-Down &  Client-Server & Centralized & Static  & Bare Metal & Context \tabularnewline\hdashline 
\cite{9} & Optimization &   &   &  Cluster  & Centralized  & Static  &  VM  &  Profit\tabularnewline\hdashline 
\cite{10} &   &   & Bottom-Up  &   &  Broker &  Dynamic & VM  & Context \tabularnewline\hdashline 
\cite{11} &  Priority  &  Predictive &   &   &   &  Dynamic  & Bare Metal  &  Context, Energy \tabularnewline\hdashline 
\cite{12} &  Optimization &   &   & Hierarchy  &   &  Event-driven &  VM  &  Cost \tabularnewline\hdashline 
\cite{13} &  Priority  &   &  Bottom-Up  &  Client-Server  &  Broker &  Dynamic  &  VM  &  QoE, Resource\tabularnewline\hdashline 
\cite{14} &  Optimization &  On Demand  &  Hybrid &  Client-Server  &  End Device, Broker &    &  VM  & Time, Resource\tabularnewline\hdashline 
\cite{115} &  Optimization & Profiling  &   & Hierarchy  &  Broker &   & Bare Metal  &  Cost \tabularnewline \hdashline
\cite{119} &   &  Predictive &   & Hierarchy  &   &  Dynamic  &  VM  &   Resource \tabularnewline\hdashline 
\cite{129} &   &   &  Hybrid &   &  Broker &  Event-driven &  Container &  Energy, Context \tabularnewline\hdashline 
\cite{21} &  Optimization &   &   & Hierarchy  &   &  Event-driven & Bare Metal  & Time \tabularnewline\hdashline 
\cite{22} &  Priority  &  On Demand  &   &  Master-Slave &  Broker &   & Bare Metal  & Time, QoE, Energy \tabularnewline\hdashline 
\cite{23} &  Optimization &  Predictive &   &   &   &  Event-driven & Bare Metal  &  Resource \tabularnewline\hdashline 
\cite{24} & Trade-off  &   &   &  Cluster  &   &   & Bare Metal  & Time, Cost \tabularnewline\hdashline 
\cite{25} &  Priority  & Profiling  &   & Hierarchy  &   & Static  & Bare Metal  & Time \tabularnewline\hdashline 
\cite{29} & Optimization &  &   &   &  Broker &  Event-driven &  VM  &  QoE, Resource, Mobility\tabularnewline\hdashline 
\cite{31} &  Optimization &   &  Bottom-Up  &  Client-Server  &  End Device  & Static  & Bare Metal  & Time, Energy \tabularnewline\hdashline 
\cite{32} &  Priority  &   &   &  Cluster  & Centralized  & Static  &  VM  & Time, Cost \tabularnewline\hdashline 
\cite{116} &  Priority  &   &  Bottom-Up  & Hierarchy  &  End Device  &  Event-driven & Bare Metal  & Time, Mobility \tabularnewline\hdashline 
\cite{33} &   &  Predictive &   & Hierarchy  & Centralized  &  Event-driven & Bare Metal  &  Context \tabularnewline\hdashline 
\cite{125} &  Priority  &   &   & Hierarchy  & Broker & Static  &   & Time, Cost \tabularnewline\hdashline 
\cite{34} &  Optimization &   &   & Hierarchy  &   & Static  &   & Time, Resource \tabularnewline\hdashline 
\cite{35} &   &   &  & Hierarchy  &   & Static  & Bare Metal  & Time, Energy \tabularnewline\hdashline 
\cite{15} &  Optimization &   &   &  Client-Server  &  End Device  & Static  &   & Time, Cost\tabularnewline\hdashline 
\cite{16} &  Optimization &   &   & Hierarchy  &   &   & Bare Metal  & Time, Energy \tabularnewline\hdashline 
\cite{17} &  Optimization &   &  Bottom-Up  &  Client-Server  &  Broker & Static  &  & Time, Energy\tabularnewline\hdashline 
\cite{18} &  Optimization & Profiling  &  Bottom-Up  &  Client-Server  &   & Static  & Bare Metal  & Time \tabularnewline\hdashline 
\cite{19} &   &   &  & Hierarchy  &  Broker & Event-driven & Bare Metal  & Time, Resource\tabularnewline\hdashline 
\cite{20} & Trade-off  &   &  Bottom-Up  &  Master-Slave &  End Device  &   &  Container &  Resource \tabularnewline\hdashline 
\cite{26} &  Optimization &   &  Hybrid &   & Centralized  & Dynamic  & VM  & Time, Profit\tabularnewline\hdashline 
\cite{27} & Trade-off  &  Predictive &   &  Client-Server  &  End Device  &   &   &  Context, Energy, Resource \tabularnewline\hdashline 
\cite{121} &  Priority   &  On Demand  &   &  Cluster  & Centralized  &   &   &  Resource \tabularnewline\hdashline 
\cite{30} &  Optimization &   &   & Hierarchy  &  Broker & Event-driven &  VM  &  Mobility \tabularnewline\hdashline 
\cite{36} &   &   &   & Hierarchy  &   &  Dynamic  & Bare Metal  & Time, Context \tabularnewline\hdashline 
\cite{126} &   &  Predictive &   & Hierarchy  &  Broker &   & VM  & Time, Cost \tabularnewline\hdashline 
\cite{131} &   &   &   & Hierarchy  &  Centralized  &   & Bare Metal  & Time, Mobility\tabularnewline\hdashline 
\cite{38} &  Priority  &  Predictive &   & Hierarchy  &  End Device  &  Event-driven &   &  Context \tabularnewline\hdashline 
\cite{108} & Trade-off  &   &   &  Cluster  &  Broker &   &  VM  & Time, Cost, Energy \tabularnewline\hdashline 
\cite{39} &   &   &   &  Master-Slave &  Broker &   & Bare Metal  &  Cost, Resource \tabularnewline\hdashline 
\cite{133} &  Optimization &  On Demand  &   &   &   & Dynamic  &  Container &  Mobility\tabularnewline\hdashline 
\cite{40} &  Optimization &   &   & Hierarchy  &   &   & Bare Metal  & Time \tabularnewline\hdashline 
\cite{118} & Priority &   &  Hybrid & Hierarchy  &  Broker &   &    & Time, Energy \tabularnewline\hdashline 
\cite{41} &  Priority  &   & Bottom-Up  &  Client-Server  &  Broker &  Dynamic  & Bare Metal  & Time, Energy \tabularnewline\hdashline 
\cite{42} &  Optimization &   &  Bottom-Up  &  Client-Server  &  End Device  &  Dynamic  & Bare Metal  & Time \tabularnewline\hdashline 
\cite{43} &  Optimization &   &   & Hierarchy  &   & Static  &  VM  & Time, Cost \tabularnewline\hdashline 
\cite{44} &  Optimization &   &   &  Cluster  &  & Static  & Bare Metal  & Time, Cost \tabularnewline\hdashline 
\cite{45} &  Optimization & Profiling  &   & Hierarchy &  Broker & Static  & Bare Metal  & Time, Cost \tabularnewline\hdashline 
\cite{46} &  Optimization &   &   &  Cluster  &  End Device  &  Dynamic  & Bare Metal  &  Cost, Energy \tabularnewline\hdashline 
\cite{48} &  Optimization & Profiling  & Top-Down  &   & Centralized &  Event-driven &   & Time \tabularnewline\hdashline 
\cite{49} &   &  On Demand  &   &   &  Broker &   &  Container & Cost \tabularnewline\hdashline 
\cite{50} &  Optimization &   &  Hybrid &  Cluster  &  End Device  &  Dynamic  & Bare Metal  & Time \tabularnewline\hdashline 
\cite{51} &  Priority  &   &   &  Cluster  &   &  Dynamic  & Bare Metal  &  Resource \tabularnewline\hdashline 
\cite{52} &  Optimization &   & Top-Down  & Hierarchy  & Centralized  & Static  &   &  Cost \tabularnewline\hdashline 
\cite{53} &  Priority  &   &  Hybrid &  Cluster  &  End Device  &  Event-driven &  &  Resource \tabularnewline\hdashline 
\cite{134} & Trade-off &   &   &  Hierarchy  & End Device &    & Bare Metal &  Time, Energy\tabularnewline \hline 
\cite{54} &  &   &  Bottom-Up  & Hierarchy  &  End Device  & Static  &  VM  & Time \tabularnewline\hdashline 
\cite{55} &  Optimization &   &  Hybrid &  Cluster  &  End Device  & Static  & Bare Metal  & Time \tabularnewline\hdashline 
\cite{56} &  Optimization &   &  Bottom-Up  &   &  End Device  &   & Bare Metal  & Time, Cost, Energy\tabularnewline\hdashline 
\cite{57} &   &   &  & Hierarchy  &  End Device  &   &  Container & Time, Resource \tabularnewline\hdashline 
\cite{58} &   &   &  & Hierarchy  &  End Device  &   &  VM & Energy \tabularnewline\hdashline
\cite{Latency} &  Priority  &   &  Hybrid & Hierarchy, Cluster &  End Device  & Static  & Bare Metal  & Time, Resource\tabularnewline\hdashline 
\cite{Context} &  Optimization &   &   &  Cluster  & Centralized, Broker & Static  & VM, Container & Context, Resource \tabularnewline\hdashline 
\cite{Edge} &  Priority  &   & Hybrid & Hierarchy, Cluster  &  Broker &  Dynamic  &  VM, Container & Time, Context \tabularnewline\hdashline 
\cite{QoE} &  Optimization & Profiling  &   & Hierarchy  &  Broker & Static & Bare Metal  &  QoE \tabularnewline\hdashline 
\cite{Profit} &  Priority  &   &  Hybrid &  Cluster  &  Broker &  Dynamic  &  VM, Container &  Profit, Cost \tabularnewline\hdashline 
\cite{60} &   &   &  Bottom-Up &  Client-Server  &  End Device  &  Dynamic  & VM, Container & Resource, Mobility\tabularnewline\hdashline 
\cite{61} &  Priority  &   &   &  Client-Server  &  End Device  &  Dynamic  &   & Time \tabularnewline\hdashline 
\cite{123} & Optimization  &   & Top-Down &  & Centralized &   &  Container &  Context \tabularnewline\hdashline 
\cite{62} &   &   &   &  Client-Server  &  End Device  &   & Bare Metal &  Context \tabularnewline\hdashline 
\cite{64} & Trade-off  &   &   &  Client-Server  & Centralized  &  Dynamic  & VM  & Time, Energy \tabularnewline\hdashline 
\cite{124} & Optimization &    &   & Hierarchy  & Broker &    &   &  Time, Energy \tabularnewline\hdashline 
\cite{120} &  Optimization &   &   &   & End Device & Static  &  VM, Container &  Energy, Resource \tabularnewline\hdashline 
\cite{68} &  Optimization &   &   &  Client-Server  &   &   & Bare Metal  & Time \tabularnewline\hdashline 
\cite{69} &  Priority  & Profiling  &   &  Client-Server  &  Broker &   &  VM  & Time \tabularnewline\hdashline 
\cite{70} & Trade-off  &   &  Bottom-Up & Hierarchy  &  Broker &   & Bare Metal  & Time, Cost, Energy \tabularnewline\hdashline 
\cite{67} &   &   &   &  Master-Slave & Centralized  &    & Bare Metal  &  Energy \tabularnewline\hdashline 
\cite{71} &  Priority  &  &   & Hierarchy  &  End Device  &   &  VM  &  Energy \tabularnewline\hdashline 
\cite{72} &  Optimization &   &   &  Client-Server  & Centralized  &  Dynamic  & Bare Metal & Mobility\tabularnewline\hdashline 
\cite{73} &   &   &  Hybrid &  Cluster  &  End Device  &   & Bare Metal  &  \tabularnewline\hdashline 
\cite{77} &   &   &  &   & Centralized  &  Event-driven &  Container &  Mobility\tabularnewline\hdashline 
\cite{78} &  Priority  &   &   & Hierarchy  &  End Device  &   &  & Resource \tabularnewline\hdashline 
\cite{128} & Optimization &   & Bottom-Up  &  Master-Slave &  Broker &   & Bare Metal &  Cost \tabularnewline\hdashline 
\cite{79} &  Priority  &   &   &  Master-Slave & Centralized  &   &  Container &  Resource \tabularnewline\hdashline 
\cite{80} &  Priority  &   &   & Hierarchy  &   &   &  Container &  Mobility\tabularnewline\hdashline 
\cite{59} & Optimization &   &  Bottom-Up  & Hierarchy &  End Device  &  Dynamic &  VM  &  QoE \tabularnewline\hdashline 
\cite{122} &  Priority  &  &   & Hierarchy, Cluster  &  Broker &   & Bare Metal  & Time, Resource \tabularnewline\hdashline 
\cite{117} &  Priority  &   &   &  Cluster  &  End Device  &   & Bare Metal  &  Context \tabularnewline\hdashline 
\cite{132} &  Priority  &   & Top-Down  &   & Centralized  &   &  VM  &  Energy, Resource \tabularnewline\hdashline 
\cite{81} &  Priority  &  &   & Hierarchy  &   &   & Bare Metal  &  Resource \tabularnewline\hdashline 
\cite{82} &  Priority  &   &   &  Cluster  &  Broker &   &  VM, Container  & Time, Resource \tabularnewline\hdashline 
\cite{83} &  Optimization &   &   &  Cluster  &  End Device  &   &   & Time, Cost \tabularnewline\hdashline 
\cite{84} & Optimization &   &   & Hierarchy  &  Broker &  Dynamic  &   & Time, Energy \tabularnewline\hdashline 
\cite{85} &  Priority  &   &   &  Cluster  &   &   & Bare Metal  &  Cost \tabularnewline\hdashline 
\cite{89} &  Optimization & Profiling  &   &  Cluster  &  Broker &   &   & Time, Resource \tabularnewline\hdashline 
\cite{90} &   & Profiling  &   &  Cluster  &  Broker &  Dynamic &   & Time, Resource \tabularnewline\hdashline 
\cite{63} &  Optimization &   &  Hybrid & Hierarchy, Cluster  &  End Device  &   &  Container &  Energy \tabularnewline\hdashline 
\cite{87} &  Optimization &   &   &  Cluster  &   &   & Bare Metal  & Time \tabularnewline\hdashline 
\cite{86} &  Optimization &   &  Bottom-Up &   & Centralized  &   & Bare Metal  & Time, Energy \tabularnewline\hdashline 
\cite{FogBus} &   &  On Demand  &  Bottom-Up  &  Master-Slave &  Broker &  Dynamic  & Bare Metal  &  \tabularnewline\hdashline 
\cite{88} & Priority &   &   &   &  Broker &   &  VM, Container &  Resource \tabularnewline\hdashline 
\cite{91} &  Optimization &   &  Bottom-Up  & Hierarchy  & Centralized  &  Event-driven &   &  Profit, Mobility\tabularnewline\hdashline 
\cite{92} &   &   &  Bottom-Up  &  Client-Server  &  Broker &  Event-driven &   & Time, Cost \tabularnewline\hdashline 
\cite{93} &  Optimization &   &   & Hierarchy  &  Broker &   &   & Time, Context \tabularnewline\hdashline 
\cite{94} & Trade-off  &   &  Hybrid &  Client-Server  &  End Device  &   &   & Time, Energy \tabularnewline\hdashline 
\cite{95} &   &   &  Bottom-Up  &  Client-Server  &  Broker &   & Bare Metal  &  \tabularnewline\hdashline 
\cite{96} &   &   &   &  Master-Slave &  End Device  &  Event-driven &  Container &  Resource \tabularnewline\hdashline 
\cite{97} &  Priority  &   &   & Hierarchy  &   & Static  & Bare Metal  & Time \tabularnewline\hdashline 
\cite{98} &   &   & Top-Down  &  Client-Server  & Centralized  &  Dynamic  &  Container &  \tabularnewline\hdashline 
\cite{99} & Trade-off  &   &  Hybrid &   &  End Device  & Static  & Bare Metal  & Time, Energy \tabularnewline\hdashline 
\cite{100} &  Optimization &  On Demand  &   &  Client-Server  &   &   &  VM  & Time, Cost \tabularnewline\hdashline 
\cite{101} &  Optimization &   &   &  Client-Server  &  End Device  &   &  Container & Time \tabularnewline\hdashline 
\cite{102} &  Priority  &   &  Hybrid &   & Centralized  & Static  &   &  Context, Resource \tabularnewline\hdashline 
\cite{103} &   &   &   &  Cluster  &   & Static  & Bare Metal  &  Resource \tabularnewline\hdashline 
\cite{104} & Optimization  &   &   &  Cluster &   &   & Bare Metal  &  Energy, Resource \tabularnewline\hdashline 
\cite{105} &   &   & Hybrid  &  Client-Server  &   &   & Bare Metal  & Time, Cost, Energy \tabularnewline\hdashline 
\cite{106} &  Priority  &  Predictive &   &  Cluster  & Centralized  &  Dynamic  &   &  Energy \tabularnewline\hdashline 
\cite{107} &  Optimization &   &   &  Client-Server  &  End Device  &  Dynamic  & Bare Metal  & Time \tabularnewline\hdashline 
\cite{109} &  Optimization &   &   &  Cluster  &   &  Event-driven &   &  Mobility\tabularnewline\hdashline 
\cite{110} &  Optimization &   & Top-Down  & Hierarchy  & Centralized  &  Dynamic  & Bare Metal  &  Resource \tabularnewline\hdashline 
\cite{127} &   &   &  Hybrid & Hierarchy  &  End Device  &   & Bare Metal  &  Energy \tabularnewline\hdashline 
\cite{111} &  Optimization &   &  Hybrid &  Client-Server  &  End Device  &   & Bare Metal  &  Energy \tabularnewline\hdashline 
\cite{112} &  Optimization &   & Top-Down  &  Client-Server  & Centralized  &   &   & Time, Cost \tabularnewline\hdashline 
\cite{113} &  Priority  &   & Top-Down  &  Master-Slave & Centralized  &  Dynamic  &  &  Profit \tabularnewline\hdashline 
\cite{114} & Trade-off  & Profiling  &  Hybrid &  Client-Server  &  Broker &  Event-driven &  Container & Time, Context, Mobility\tabularnewline\hdashline 
\addlinespace[2ex]
\caption{Summary of existing application placement techniques}\label{Tab:related_appPlacement} 
\end{longtable}}}
\end{landscape}
\par 3. Most of the existing works prefer Cloud to place applications when there is no resource available in the corresponding Fog infrastructure \cite{Profit}. However, they have not discussed the confederation of Fog infrastructure owned by different service providers. As a consequence, the scope of performance improvement lessens and the providers fail to harness the monetary benefits.
\par 4. The consolidation of Fog nodes can save energy. However, this operation can alter the topology and orientation of Fog resources and affect the collaborative execution of applications. Despite of such impact, current researches barely look into this issue \cite{29}. 
\section{Application Maintenance} \label{Sec_Maintenance}
\begin{figure}[!t]
\centering 
\includegraphics[width=140mm, height= 40mm]{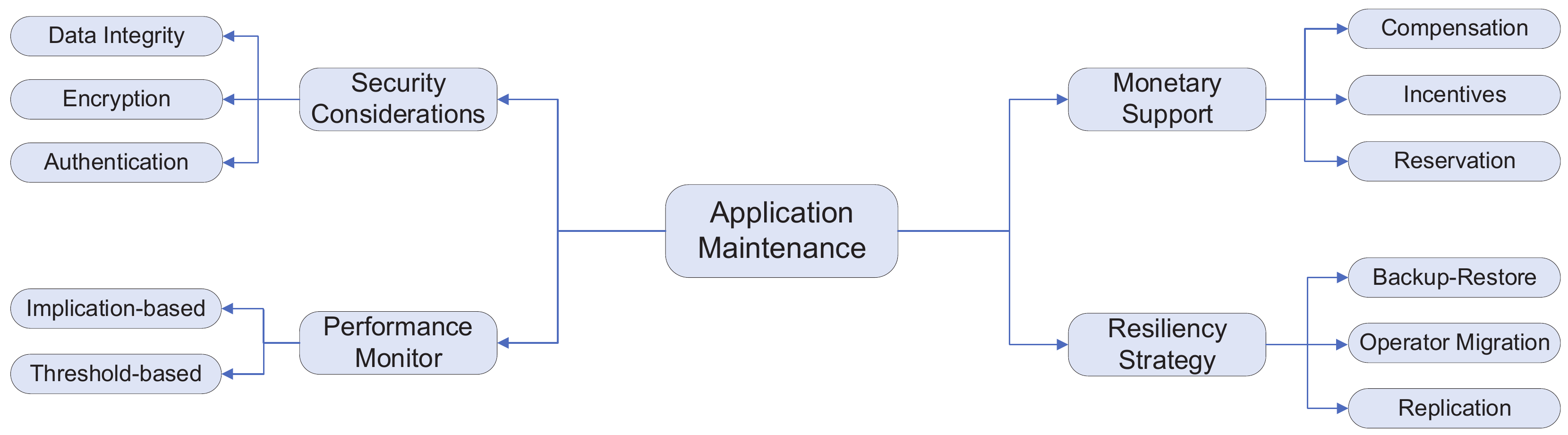}
\caption{Taxonomy on application maintenance}
\label{Fig:maintenance}
\end{figure}

Robust application maintenance operations are required to secure the access of all legitimate users to the application outcomes in Fog environments. Additionally, if these operations are conducted in both proactive and reactive manner, the uncertain failure of Fog nodes and the performance degradation of applications can be mitigated to a certain extent. The demand of application maintenance can also trigger the necessity to introduce additional features such as check pointing and partitioning to the functional layout of application architecture. Moreover, the requirements of operator migration can initiate the event-driven application placement. Fig. \ref{Fig:maintenance} illustrates a taxonomy of different elements of application maintenance. In following subsections, they are discussed in detail.
%
%
\subsection{Security Considerations}
Fog computing functions at the edge network. The attackers can access the Fog infrastructure easily and resist the smooth execution of applications by generating security threats including information impairment, identity disclosure, replay and denial of service (DoS) attacks. Therefore, Fog infrastructure is required to specify the security while executing the applications. Three different types of security are widely considered in Fog computing. 
\subsubsection{Data Integrity:}
In some cases, various sensitive data and their processing outcomes; for instance, the electronic health reports are consistently analysed by different parties including the hospitals and insurance companies. Fog computing offers easy access to these data and information with a guarantee of no alteration. Such initiatives help data integrity of applications in Fog environments. \cite{FogBus}, \cite{78} and \cite{80} discuss data integrity for Fog computing.
\subsubsection{Encryption:}
In Fog computing, extensive data exchange operations are conducted between the IoT devices, Fog nodes and Cloud datacentres. Encryption not only hides the details of the transferred data, but also protects the credentials of legitimate users to access the Fog resources. In \cite{8}, \cite{28}, \cite{33}, \cite{73} and \cite{83}, encryption for Fog computing is discussed. 
\subsubsection{Authentication:}
Authentication helps to identify the legitimate user of application services and Fog resources. Sometimes, authentication is robustly applied at the receiver side to control the data access rate of different entities. \cite{33}, \cite{57}, \cite{63}, \cite{78} and \cite{98} consider authentication as a security measure for application maintenance. 
\subsection{Performance Monitor}
During application runtime, a continuous monitoring of resources is required to maintain the execution flow of the applications at the desired level. Two different types of performance monitoring techniques are widely used in Fog computing. 
\subsubsection{Implication-based:}
In this technique, the current synthesis of application and resources are implied to predict the future performance trends. Implication-based performance monitoring assists in maintaining the execution of the applications in a proactive manner. \cite{4}, \cite{14}, \cite{23}, \cite{54} and \cite{91} discuss implication-based performance monitoring for Fog.  
%
%
%
\subsubsection{Threshold-based:}
When an application is executed in a Fog node, its key performance indicator such as processor usage and memory consumption are compared with a dynamically set or predefined threshold value. If the state of the indicator does not match with the threshold, necessary decisions are taken to continue the execution of the application in Fog environments. Unlike the implications, this approach helps in reactive application maintenance. \cite{13}, \cite{18}, \cite{19}, \cite{34} and  \cite{88} discuss about threshold based performance monitoring. 
%
\subsection{Monetary Support}
Monetary support defines how Fog service providers nurture the economic interests of the users while executing their requested applications. At the same time, the monetary support helps providers to attain the trust of the users that controls the relinquish rate. Three types of monetary support are found in Fog computing.
\subsubsection{Compensation:}
Due to some uncertain events such as node failure and security threats, the service level agreement between the users and the providers can be violated. In these cases, provider offers compensation to users so that they can rely on the Fog based execution of their requested applications. Compensation also helps providers in quantifying the aim of reducing service violations. \cite{Profit}, \cite{compensation1}, \cite{compensation2} and \cite{49} highlight compensation approach for Fog computing. 
\subsubsection{Incentives:}
At the edge network, there exist different idle computing resources that can be used as potential Fog nodes. Fog service providers often harness them to meet the additional demand of users by providing incentives to the owner of the resources. In some cases, for relaxing the stringent requirements, the users also receive incentives from the service providers. \cite{39}, \cite{42}, \cite{52}, \cite{105} and \cite{113} consider incentives as monetary support to maintain application execution in Fog. 
\subsubsection{Reservation:}
Reservation assists users to provision a certain number of applications at any given time on fixed charges despite of the current load on the Fog infrastructure. In this case, Fog computation is referred as a subscription-based utility. \cite{36}, \cite{reservation1}, \cite{reservation2} and \cite{Edge} discuss reservation as a monetary support for the application users in Fog computing. 
\subsection{Resiliency Strategy}
Resiliency strategy denotes how Fog computing continues application execution after the occurrences of uncertain events and failures. Resiliency strategies assist in enhancing the reliability of the system. Three types of resilience strategies are widely adopted in Fog computing. 
\subsubsection{Backup-Restore:}
In this strategy, the intermediate results and different execution phase of an application are continuously stored so that the execution of the application can be re-initiated soon after the anomaly occurs. This operation is performed by setting some check points and temporary storage operations during the application runtime. Backup-restore is feasible when a one-to-one interaction happens between the users and the applications. \cite{18}, \cite{65}, \cite{73} and \cite{75} discuss backup-restore as a resiliency strategy. 
\subsubsection{Replication:}
In this resiliency strategy, multiple instances of an application are run across different Fog nodes. Replication ensures the satisfaction of user requests through an application even after the failure of its several instances. Unlike backup-restore, replication is well-suited for supporting the one-to-many interactions between the users and the applications. Replication for Fog is discussed in \cite{10}, \cite{19}, \cite{20}, \cite{34} and \cite{92}. 
\subsubsection{Operator Migration:}
In case of node failure or mobility of the requesting entities, the execution of applications is often shifted from one node to another Fog node. Such migration of operators happens dynamically so that application execution continues without interrupting the user experience. As a resiliency strategy, operator migration differs from the backup-restore and replication because of its ease of scalability. \cite{29}, \cite{30}, \cite{38}, \cite{69} and \cite{98} discuss operator migration in Fog. 
\subsection{Research Gaps in Application Maintenance} 
\par Table \ref{Tab:related_appMaintain} summarizes the existing application maintenance operations in Fog computing. Although extensive research initiatives have been taken, there are still some issues that require to be investigated for efficient application maintenance in Fog computing. They are discussed below:
\par 1. In many research works, compute intensive algorithms are used to secure the data transmission within Fog environments \cite{PaolaPaper}. However, they have not considered that heavyweight security techniques slow down the legitimate access to application services and resources.
\par 2. Streaming applications require reserved resources so that their processing destinations do not change very frequently. However, while making such arrangements, the existing works barely consider the waiting time of the other less-interactive applications \cite{reservation1}.
\par 3. There are a good number of research works that mention check pointing and replication as the means of fault tolerance in Fog computing \cite{74}. However, to deal with the scarcity of resources, they have not provided any concrete model that can dynamically tune the frequency of check points within an application and change the number of replications in Fog environments. 
\newpage
{\scriptsize
\begin{longtable}{|>{\raggedright}p{3.5 cm}| >{\raggedright} p{1.9 cm} >{\raggedright} p{2.2 cm} >{\raggedright} p{1.9cm} >{\raggedright} p{2 cm}|}\hline
Works  & \multicolumn{4} {c|}{Application Maintenance} \tabularnewline \cline{2-5}  
& Security Considerations & Performance Monitor & Monetary Support & Resiliency Strategy\tabularnewline\hline 
\cite{reservation1} &   &    & Reservation  &     \tabularnewline\hdashline 
\cite{4} &   & Implication   &   &   \tabularnewline\hdashline 
\cite{8} &  Encryption  &    &   &   \tabularnewline\hdashline 
\cite{10} &   &    &   &  Replication \tabularnewline\hdashline 
\cite{13} &   &  Threshold  &   &   \tabularnewline\hdashline 
\cite{14} &   & Implication   &   &   \tabularnewline\hdashline 
\cite{23} &   & Implication   &   &   \tabularnewline\hdashline 
\cite{29} &   &    &   &  Migration  \tabularnewline\hdashline 
\cite{34} &   &  Threshold  &   & Replication \tabularnewline\hdashline 
\cite{18} &   &  Threshold  &   & Backup-Restore  \tabularnewline\hdashline 
\cite{19} &   &  Threshold  &   &  Replication \tabularnewline\hdashline 
\cite{20} &   &    &   &  Replication \tabularnewline\hdashline 
\cite{28} &  Encryption  &    &   &   \tabularnewline\hdashline 
\cite{30} &   &    &   &  Migration  \tabularnewline\hdashline  
\cite{36} &   &  Threshold  & Reservation  &   \tabularnewline\hdashline 
\cite{37} &   &  Threshold  &   &   \tabularnewline\hdashline 
\cite{38} &   &    &   &  Migration  \tabularnewline\hdashline 
\cite{39} &   &    & Incentives  &     \tabularnewline\hdashline 
\cite{40} &   &     &   & Replication \tabularnewline\hdashline 
\cite{reservation2} &   &    & Reservation  &     \tabularnewline\hdashline 
\cite{42} &   &    & Incentives  &     \tabularnewline\hdashline 
\cite{46} &   &  Threshold  &   &   \tabularnewline\hdashline 
\cite{48} &   &  Threshold  &   &   \tabularnewline\hdashline 
\cite{49} &   &    & Compensation  &     \tabularnewline\hdashline 
\cite{50} &   &  Threshold  &   &   \tabularnewline\hdashline 
\cite{52} &   &    & Incentives  &     \tabularnewline\hdashline 
\cite{53} &   &  Threshold  &   &   \tabularnewline\hdashline 
\cite{54} &   &  Implication  &   &   \tabularnewline\hdashline 
\cite{57} & Authentication &  Threshold  &   &   \tabularnewline\hdashline 
\cite{Latency} &    &     &   & Migration  \tabularnewline\hdashline 
\cite{Profit} &    &    & Compensation  &   \tabularnewline\hdashline
\cite{62} &   &  &   & Backup-Restore  \tabularnewline\hdashline 
\cite{65} &   &  &   & Backup-Restore  \tabularnewline\hdashline 
\cite{66} &   &  Threshold  &   &   \tabularnewline\hdashline 
\cite{69} &   &  Implication  &   & Migration  \tabularnewline\hdashline 
\cite{73} & Encryption &    &   & Backup-Restore  \tabularnewline\hdashline 
\cite{74} &   &    &   & Backup-Restore, Replication \tabularnewline\hdashline 
\cite{75} &   &  Threshold &   & Backup-Restore  \tabularnewline\hdashline 
\cite{76} &   &  Implication, Threshold &   &    \tabularnewline\hdashline 
\cite{77} &   &     &   & Migration  \tabularnewline\hdashline 
\cite{78} & Integrity, Authentication &    &   &   \tabularnewline\hdashline 
\cite{80} & Integrity &    &   & Migration  \tabularnewline\hdashline 
\cite{83} & Encryption &    &   & Replication  \tabularnewline\hdashline 
\cite{FogBus} & Integrity &    &    &   \tabularnewline\hdashline
\cite{88} &   &  Threshold  &   &   \tabularnewline\hdashline 
\cite{90} &   &     &   & Migration  \tabularnewline\hdashline 
\cite{63} & Authentication &    &   &   \tabularnewline\hdashline 
\cite{91} &   &  Implication  &   & Migration   \tabularnewline\hdashline 
\cite{92} &   &  Threshold  &   & Replication   \tabularnewline\hdashline 
\cite{93} & Authentication  &     &   &   \tabularnewline\hdashline 
\cite{95} & Authentication  &     &   &   \tabularnewline\hdashline 
\cite{98} & Authentication  &     &   & Migration  \tabularnewline\hdashline 
\cite{compensation2} &    &    & Compensation  &   \tabularnewline\hdashline
\cite{104} &   &    &   & Replication   \tabularnewline\hdashline 
\cite{105} &   &    & Incentives  &     \tabularnewline\hdashline 
\cite{compensation1} &    &    & Compensation  &   \tabularnewline\hdashline
\cite{109} &   &    &   & Migration    \tabularnewline\hdashline 
\cite{113} &   &    & Incentives  &     \tabularnewline\hdashline 
\cite{114} &   &    &   & Migration    \tabularnewline\hline 
\addlinespace[2ex]
\caption{Summary of existing application maintenance operations} \label{Tab:related_appMaintain} 
%
%
\end{longtable}}
%
%
\section{A Perspective Application Management Framework for Fog} \label{Sec_Model}
In some cases, the existing application management strategies for Fog computing can be used to solve the classical research problems in Mobile distributed computing. \cite{MDC}. This perception is often used to bind the concept of Mobile Cloud computing (MCC), Mobile Edge computing (MEC) and Fog computing to a single body. However, these computing paradigms differ from each other in both architecture and operations. For example, MCC facilitates users to offload the compute and data intensive mobile applications to Cloud for execution and overcomes the limitations of smart phones in terms of energy, storage and computation. MCC is designed with an additional computing layer of Cloudlets in between the smart phones and Cloud datacentres to offer latency-sensitive mobile application services \cite{MCCDef}. Conversely, in MEC, a virtualized server is deployed at the cellular base station to ensure flexible and rapid initiation of cellular services for the users. MEC offers real-time access to radio network information to endorse Tactile Internet, interactive gaming and virtual reality applications through 5G \cite{5G-Ritu}. In comparison to MCC and MEC, Fog computing mostly deals with the IoT-drievn use cases at the edge network. Rather than harnessing virtual cellular base stations or Cloudlets, Fog aims at building a Cloud of Things in the proximity through dedicated or ad-hoc networking. In Fog, the computing infrastructure is multi-tiered, whereas for MEC and MCC, it is 2 and 3-tiered respectively \cite{Levels}. Most importantly, Fog computing provides a scope to distribute the application management operations across different tiers of the computing infrastructure, however, in other paradigms, this scope is very limited. To illustrate this feature of Fog, a perspective framework is depicted in Fig. \ref{Fig:model}. At each infrastructure level, the components of this framework performs some specific operations related to application management. They are summarized below:
\begin{figure}[!h]
\centering 
\includegraphics[width=90mm, height= 90mm]{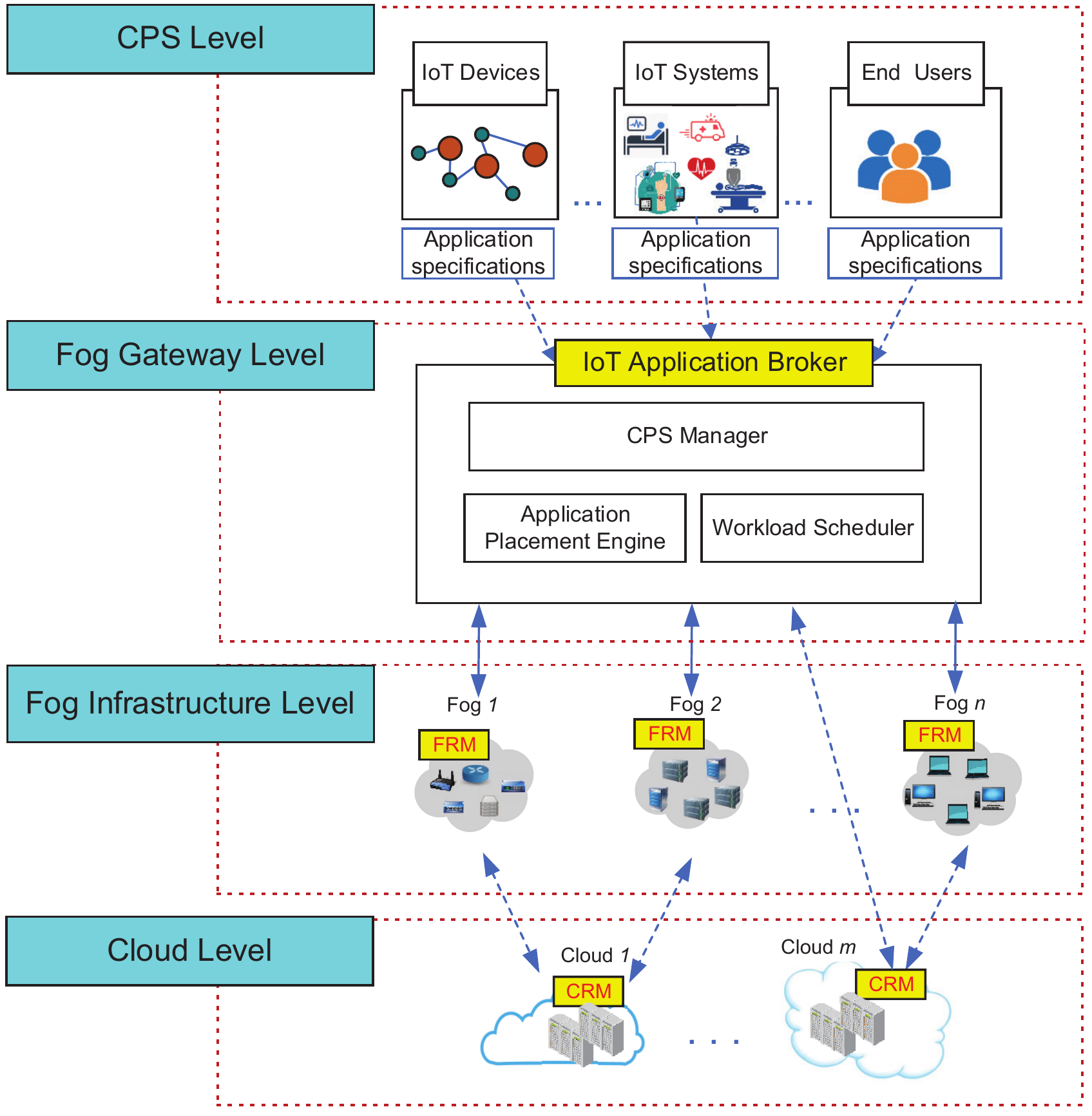}
\caption{A perspective model for application management in Fog}
\label{Fig:model}
\end{figure}
\subsection{Cyber Physical System Level}
At this level, IoT-enabled CPSs reside that request the Fog environment for the services of different applications. In this case, the IoT Application Brokers (IABs) deployed at the Fog gateway level assist the CPSs. While making the requests, the CPSs forward the specification of the applications including the workload type, the frequency of incoming data, the form of application services and the QoS requirements such as service deadline, budget and user expectations to the IABs. Additionally, the CPSs can host some components of the requested applications for data pre-processing.  
\subsection{Fog Gateway Level}
In Fog Gateway Level, an IAB consists of CPS Manager, Application Placement Engine (APE) and Workload Scheduler. The CPS manager contains the meta-data of multiple versions of an application having different programming models, functional layouts and interaction methods. The CPS manager also interacts with the APE to get the state of resources such as their orientation and type within the Fog infrastructure and Cloud level. Based on the accumulated information from different levels, the CPS manager determines the most suitable architecture of the applications for placement. Later, the APE estimates the resources of executing the application, identifies the placement metrics as per the application QoS requirements of the CPSs and sets the mapping technique accordingly. To place the applications physically over the resources, the APE communicates with the Fog Resource Manager (FRM) and Cloud Resource Manager (CRM) of the Fog infrastructure and Cloud level. After placement, the Workload Scheduler finds out the feasible placement strategy to dispatch the inputs to the applications based on the dynamics of the Fog environment.
\subsection{Fog Infrastructure and Cloud Level}
The FRMs and CRMs of Fog infrastructure and Cloud datacenter store the application executables and they are responsible for allocating resources for application execution. They also monitor the status and performance of the resources and conduct application maintenance operations including service backup and replication. Additionally, they deal with the uncertain node failures, resource outage and security attacks to ensure reliability during application execution. Based on their implications, the CPSs and IABs tune the specifications of the application architecture and modify the placement approaches.
\par Nevertheless, this framework only provides a abstract view of distributing application management operations in different infrastructure levels within the Fog computing environments. This framework can also contribute to develop new policies for runtime service orchestration, multi-level resource provisioning, application execution migration and Fog standardization.
\section{Future Research Directions} \label{Sec_Future}
In this section, we discuss several future research directions that can guide the respective community to leverage the existing solutions and make further progress in the field of Fog computing. These directions are listed below.
\par \textbf{Trade-off between energy and accuracy}: The exists a complex relation between the accuracy level applications, the sensing rate of IoT devices and the energy consumptions of Fog nodes \cite{FogBus}. Based on such relation, a policy to dynamically tune the accuracy level and the sensing frequency of the IoT devices can be developed for meeting the energy constraints of the Fog nodes; especially when the renewable power sources are used.
\par \textbf{Artificial intelligence-based application management}: Currently, artificial intelligence is receiving significant attention due its ability of solving complex problems. The training data required to build an artificial intelligent system is very easy to accumulate in Fog \cite{52}. Artificial intelligence-based application management can help to predict the future resource requirements, context variation and nodal failures more precisely, and manage the applications accordingly.
\par \textbf{Pricing and detailed estimation of Fog resources}: The Cloud-based pricing models for subscription-oriented services cannot be directly applied to Fog computing due to the localized demand and distributed deployment of the IoT-enabled systems. For the same reasons, resource over-provisioning can also occur in Fog computing environments \cite{Profit}. Therefore, detailed estimation of resources in Fog computing is needed that can consider the number of IoT devices within the CPS and their future service requirements simultaneously. Such researches will also help to develop an efficient business model for the Fog computing environments.

\par \textbf{Trusted service orchestration in Fog}: The Fog infrastructure can be private or public. The publicly available Fog infrastructure is highly exposed to security threats. On the other hand, service of private-owned Fog infrastructure is subjected to lack of transparency \cite{cloudLab1}. In this case, a trusted service orchestration policy is required to ensure the collaboration and reliability between different types of Fog computing infrastructure.
\par \textbf{Fog node consolidation and scaling}: Fog nodes are resource constrained. Inclusion of more Fog nodes can alleviate this limitation. However, it increases deployment cost, communication interference and energy consumption at the edge network \cite{Ritu-FGCS}. In this case, dynamic consolidation and scaling of Fog nodes as per the computational demand can be helpful.
\par \textbf{Application-specific management}: Fog computing is developed to execute various complex IoT applications from different domains including smart healthcare, city, agriculture and industry \cite{65}. These IoT applications have specific requirements and need specialized support. Application specific management polices can be helpful in dealing with them in Fog.
\par \textbf{Task sharing and re-usability:} Applications can share a particular task among themselves to optimize the computational load on Fog nodes \cite{Demystify}. Besides, the task executables of recently terminated applications can be also re-used for other applications. To perform such operations, shared caching techniques and policies are required to be developed in the context of Fog computing.
\section{Summary and Conclusions} \label{Sec_Conclusion}
Fog computing is gradually turning into an integral component of smart systems because of its wide-spread features for supporting IoT-driven use cases. To exploit the benefits of Fog computing, the efficient management of applications over the Fog nodes is very important. In both academia and industry, numerous initiatives have already been taken to standardize the Fog computing concept for managing the IoT applications. In this work, we reviewed the existing application management strategies in Fog computing from the perspectives of application architecture, placement and maintenance. We proposed separate taxonomy for each of the aspects of application management and discussed their associated research gaps. We also highlighted a perspective model for managing applications in Fog environments and mentioned several research directions for further improvement of Fog computing.
\bibliographystyle{ACM-Reference-Format}
\bibliography{ApplicationManagement}

\end{document}